\documentclass[twoside,english]{elsarticle}
\usepackage[T1]{fontenc}
\usepackage[latin9]{inputenc}
\usepackage{geometry}
\usepackage{units}
\usepackage{graphicx}
\usepackage{setspace}
\makeatletter

\providecommand{\tabularnewline}{\\}

\journal{Journal of Theoretical Biology}

\usepackage{savesym}
\savesymbol{eqref}
\usepackage[version=3]{mhchem}
\restoresymbol{dmb}{eqref}
\usepackage{lineno} 
\usepackage{tikz}

\usetikzlibrary{shapes,arrows}

\usepackage{pgfplots}
\usepackage{units}
\usepackage{epstopdf}

\makeatother

\usepackage{babel}
\begin{document}

\title{Determination of personalized diabetes treatment plans using a two-delay
model}

\author[CUAPPM]{S.~M.~Kissler}

\ead{Stephen.Kissler@colorado.edu}

\author[JH]{C.~Cichowitz}

\ead{cody.cichowitz@jhmi.edu}

\author[CUCompSci]{S.~Sankaranarayanan}

\ead{srirams@colorado.edu}

\author[CUAPPM]{D.~M.~Bortz\corref{cor1}}

\ead{dmbortz@colorado.edu}

\ead[url]{http://mathbio.colorado.edu}

\address[CUAPPM]{Department of Applied Mathematics, University of Colorado, Boulder,
CO (USA) 80309-0526}

\address[JH]{Department of Medicine, Johns Hopkins University, Baltimore, MD (USA)
21224}

\address[CUCompSci]{Department of Computer Science, University of Colorado, Boulder,
CO (USA) 80309-0430}

\cortext[cor1]{Corresponding author}
\begin{abstract}
Diabetes cases worldwide have risen steadily over the past decades,
lending urgency to the search for more efficient, effective, and personalized
ways to treat the disease. Current treatment strategies, however,
may fail to maintain ultradian oscillations in blood glucose concentration,
an important element of a healthy alimentary system. Building upon
recent successes in mathematical modeling of the human glucose-insulin
system, we show that both food intake and insulin therapy likely demand
increasingly precise control over insulin sensitivity if oscillations
at a healthy average glucose concentration are to be maintained. We
then suggest guidelines and personalized treatment options for diabetic
patients that maintain these oscillations. We show that for a type
II diabetic, both blood glucose levels can be controlled and healthy
oscillations maintained when the patient gets an hour of daily exercise
and is placed on a combination of Metformin and sulfonylurea drugs.
We note that insulin therapy and an additional hour of exercise will
reduce the patient's need for sulfonylureas. Results of a modeling
analysis suggest that a typical type I diabetic's blood glucose levels
can be properly controlled with a constant insulin infusion between
0.45 and 0.7 $\unitfrac{\mu U}{ml\cdot min}$. Lastly, we note that
all suggested strategies rely on existing clinical techniques and
established treatment measures, and so could potentially be of immediate
use in the design of an artificial pancreas.\end{abstract}
\begin{keyword}
Artificial pancreas \sep Blood glucose \sep Insulin sensitivity
\sep Personalized medicine \sep Ultradian oscillations
\end{keyword}
\maketitle

\section{Introduction\label{sec:Introduction}}

The number of cases of diabetes in the United States has doubled since
2000 and more than tripled since 1990, with current figures estimating
about 25.8 million cases (\citet{CentersforDiseaseControlandPrevention2011,CDC2012}).
The term ``diabetes'' refers to a range of conditions, varying in
origin and severity, characterized by chronic high levels of glucose
in the blood, which can lead to peripheral neuropathy, cardiovascular
disease, blindness, and even death (\citet{ADA2003,Boulton1998,Kannel1979}).
Type I diabetes is caused by autoimmune attack on the insulin-producing
pancreatic $\beta$-cells. Its onset is largely dictated by genetic
factors, and the disease is usually present from early in life (\citet{Daneman2006}).
Type II diabetes is characterized by decreased sensitivity to insulin,
making it more difficult for cells to utilize glucose and eventually
impairing insulin secretion by pancreatic $\beta$-cells (\citet{Stumvoll2005}).
While generally less severe, Type II is also much more common and
possesses many risk factors ranging from genetics to obesity. Each
case is unique and no two people have the same ability to utilize
glucose, the same insulin production rate, or the same lifestyle.
With no known cure for diabetes, lifelong treatment is generally the
only option. It is therefore of great importance for an individual's
treatment plan to be tailor-made for his or her specific condition.

The American Diabetes Association (ADA) recommends a combination of
diet, exercise, medication, and insulin therapy to treat diabetes.
These treatments are used to lower blood glucose concentration (BGC)
to a healthy level (\citet{ADA2013}). However, another important
factor is often overlooked: blood glucose levels in non-diabetic individuals
also fluctuate by about $10$\% every two hours or so. These so-called
\emph{ultradian} \emph{oscillations} (i.e., taking place multiple
times each day) were first noted by Hansen in 1923, and various studies
since have underlined their prominence and functional importance in
regulating glucose concentration (\citet{Drozdov1995,Hansen1923,Simon1987,Simon2000}).
The root cause of these oscillations is not fully understood, though
evidence suggests that delayed feedback between insulin-producing
pancreatic $\beta$-cells and the liver may be a significant contributing
factor \citet{Li2006}. As these oscillations are natural and indicative
of healthy insulin dynamics, any effective treatment strategy should
aim to maintain these oscillations.

With these points in mind, our goal is to develop a systematic strategy
to determine a personalized treatment plan for lowering a diabetic's
BCG to within the ADA-specified range (between 70 and 130 $\unitfrac{mg}{dl}$
before meals (\citet{AmericanDiabetesAssociation2008})). This treatment
plan should retain the ultradian glucose oscillations observed in
healthy individuals and should rely on existing standard treatment
measures, i.e.\ diet, exercise, insulin therapy, and/or medication.
It should be straightforward enough to be programmed into a medical
device such as an artificial pancreas. Finally, the information necessary
to personalize the treatment plan should be readily available from
existing clinical procedures. To accomplish these goals we will study
a mathematical model of the human glucose-insulin system that explicitly
accounts for the treatment methods proposed by the ADA. We present
this model in Section \ref{sec:Model}. In Section \ref{sec:Analysis}
we identify the conditions under which a person's BGC will reach an
acceptable range and will oscillate. To illustrate how this method
can be put into practice, we perform a hypothetical case study in
Section \ref{sec:Results}, in which we set forth viable plans to
treat a Type I and a Type II diabetic. We conclude with our results
in Section \ref{sec:Conclusions} and propose areas for further research.

\section{Model Presentation\label{sec:Model}}

\begin{figure}[htbp]
\begin{centering}
\tikzstyle{input} = [ellipse, draw, fill=green!20, 
     text badly centered, node distance=1.5cm, inner sep=0pt, minimum height=2em]
\tikzstyle{substrate} = [rectangle, draw, fill=blue!20, 
    text width=5em, text centered, rounded corners, minimum height=2em]
\tikzstyle{line} = [draw, -latex']
\tikzstyle{organ} = [draw, ellipse,fill=red!20, node distance=1.5cm,
    minimum height=2em]
\tikzstyle{space} = [rectangle, draw, fill=white!20, 
    text width=5em, text centered, rounded corners, minimum height=2em]
    \begin{tikzpicture}[node distance = 2cm, auto]
    \node [space, draw=none] (glucagon) {};
    \node [substrate, left of=glucagon, node distance = 2.5cm] (glucose) {Glucose($G$)};
    \node [substrate, right of=glucagon, node distance = 2.5cm] (insulin) {Insulin($I$)};
    \node [input, left of=glucose, node distance=3.5cm] (ingestion) {Ingestion};
    \node [input, above of=insulin] (infusion) {Insulin Infusion};
    \node [organ, below of=glucagon] (pancreas) {Pancreas};
    \node [organ, above of=glucagon] (liver) {Liver};
    \node [organ, below of=pancreas] (muscle) {Muscle/Fat};
    \node [organ, left of=muscle, node distance=2.5cm] (cns) {CNS};
    \node [input, right of=insulin, node distance=3.5cm] (clearance) {Clearance};
    \path [line, dashed] (glucose) -- (pancreas);
    \path [line] (pancreas) -- (insulin); 
    \path [line, dotted] (insulin) -- (liver);
    \path [line] (liver) -- (glucose);
    \path [line] (infusion) -- (insulin); 
    \path [line] (insulin) -- (clearance);
    \path [line] (glucose) -- (muscle);
    \path [line] (glucose) -- (cns); 
    \path [line] (ingestion) -- (glucose); 
    \path [line, dashed] (insulin) |- (muscle);
    \node[draw=none] at (-1.5,1) {(Delay)};
    \node[draw=none] at (1.6,-1) {(Delay)};
    \node[draw=none] at (-4.25,0.2) {$G_{in}$};
    \node[draw=none] at (-2.8,-2) {$f_2$};
    \node[draw=none] at (-1.5,-2) {$f_3 \cdot f_4$};
    \node[draw=none] at (2.25,.8) {$I_{in}$};
    \node[draw=none] at (4.25,0.3){$\frac{V_{max}I(t)}{K_{M}+I(t)}$};
    \node[draw=none] at (1,-.6){$f_5$};
    \node[draw=none] at (-1,.6){$f_1$};
\end{tikzpicture}

\par\end{centering}

\centering{}\caption{Schematic diagram of the human glucose-insulin system. Solid lines
denote production/consumption of a substrate (glucose or insulin),
dotted lines denote inhibition by a substrate, and dashed lines denote
encouragement by a substrate. Ingested food is converted to glucose,
which the body uses to fuel biological processes. Glucose also stimulates
pancreatic $\beta$-cells to produce insulin, which in turn inhibits
the liver's production of glucose. The central nervous system (CNS)
processes glucose without insulin, whereas insulin enhances glucose
uptake by muscle and fat cells. Thus, when blood glucose levels are
high, insulin is produced to stimulate glucose uptake and to slow
the production of further glucose from the liver. When blood glucose
levels are low, insulin is produced more slowly and the liver's production
of glucose speeds up. This feedback loop helps to keep a person's
blood glucose levels in a state of oscillatory homeostasis. \label{fig:gi_schematic} }
\end{figure}
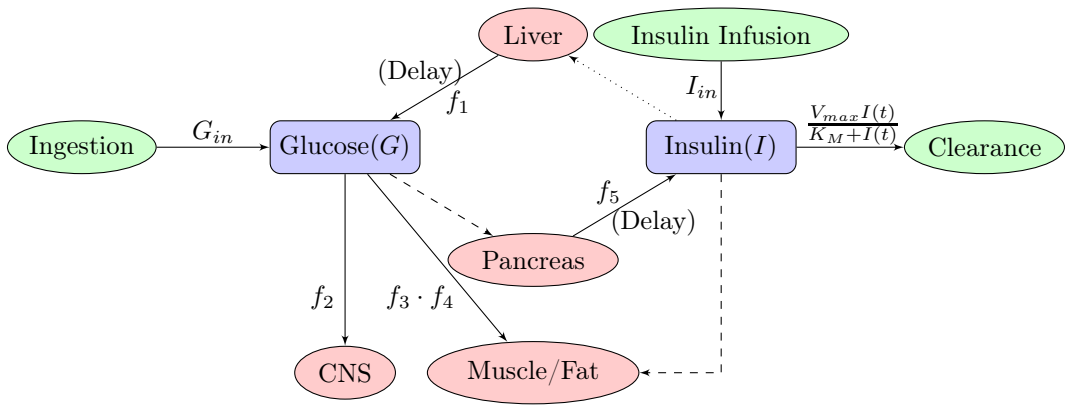

We begin with a schematic model of the human glucose-insulin system,
illustrated in Figure \ref{fig:gi_schematic}. In the human body,
ingested food is converted to glucose, which fuels bodily functions
and encourages the production of insulin. This insulin, in turn, slows
down further glucose production to prevent a buildup of glucose in
the blood stream. The model given by equations (\ref{eqn:Gmodel})
and (\ref{eqn:Imodel}) mathematically describes this process. 

\begin{align}
G' & =G_{in}+f_{1}(I(t-\tau_{2}))-f_{2}(G(t))-\gamma[1+s(m-m_{b})]f_{3}(G(t))f_{4}(I(t))\label{eqn:Gmodel}\\
I' & =I_{in}+\beta f_{5}(G(t-\tau_{1}))-\frac{V_{max}I(t)}{K_{M}+I(t)}\label{eqn:Imodel}
\end{align}

\begin{table}[b]
\caption{Parameter values for model equations (\ref{eqn:Gmodel}) and (\ref{eqn:Imodel})}

\begin{centering}
\begin{tabular}{|c|c|c|c|}
\hline 
Parameters  & Units  & Range  & Meaning \tabularnewline
\hline 
$\beta$  & $-$  & $0-1$  & Relative pancreatic $\beta$-cell function\tabularnewline
$\gamma$  & $-$  & $0-1$  & Relative insulin sensitivity\tabularnewline
$G_{in}$  & $\unitfrac{mg}{dl\cdot min}$ & $0-1.08$  & Glucose intake rate\tabularnewline
$I_{in}$  & $\unitfrac{\mu U}{ml\cdot min}$  & $0-2$  & Insulin infusion rate\tabularnewline
$K_{M}$  & $\unitfrac{\mu U}{ml}$  & 2300  & Insulin degrading enzyme's half-saturation concentration\tabularnewline
$m$ & $min$ & 0-120 & Daily minutes of physical activity\tabularnewline
$m_{b}$ & $min$ & 60 & Baseline minutes of physical activity \tabularnewline
$s$ & $1/min$ & 0.0072 & Rate of insulin sensitivy increase per minute of exercise\tabularnewline
$V_{max}$  & $\unitfrac{\mu U}{ml\cdot min}$  & 150  & Maximum insulin clearance rate\tabularnewline
\hline 
\end{tabular}
\par\end{centering}

\label{tab:model_params} 
\end{table}

\begin{table}[t]
\caption{Definitions of functions $f_{1}$ - $f_{5}$ from model equations
(\ref{eqn:Gmodel}) and (\ref{eqn:Imodel}). Parameter values are
given in Table \ref{tab:params}.}

\begin{centering}
\begin{tabular}{|l|c|}
\hline 
\multicolumn{1}{|c|}{Modeling Term} & Physiological Process \tabularnewline
\hline 
$f_{1}(I)=R_{g}/(1+\exp(\alpha(I/V_{p}-C_{5})))$  & Hepatic glucose production \tabularnewline
$f_{2}(G)=U_{b}(1-\exp(-G/(C_{2}V_{g})))$  & CNS glucose utilization\tabularnewline
$f_{3}(G)=G/(C_{3}V_{g})$  & Muscle/fat glucose utilization\tabularnewline
$f_{4}(I)=U_{0}+(U_{m}-U_{0})/(1+\exp(-\beta\ln(I/C_{4}(1/V_{i}+1/(Et_{i})))))$  & Muscle/fat insulin uptake\tabularnewline
$f_{5}(G)=R_{m}/(1+\exp((C_{1}-G/V_{g})/a_{1}))$  & Pancreatic insulin production\tabularnewline
\hline 
\end{tabular}
\par\end{centering}

\label{tab:f1f5} 
\end{table}


For clarity, let us explain the links between the terms in Equations
(\ref{eqn:Gmodel}) and (\ref{eqn:Imodel}) and the processes depicted
in Figure \ref{fig:gi_schematic}. We first note that glucose concentration
($G$) can increase via two pathways: (1) ingestion and (2) production
by the liver (commonly called \textit{hepatic production}). We first
consider ingestion, for which we represent glucose intake rate by
$G_{in}$. We make this term constant because, if it were instead
periodic (as in the case of multiple daily meals), this periodicity
would automatically induce ultradian glucose oscillations. \citet{Simon1987}
demonstrated that ultradian glucose oscillations exist in healthy
individuals even when ingesting glucose at a constant rate, and we
want to ensure that our model accounts for this behavior.

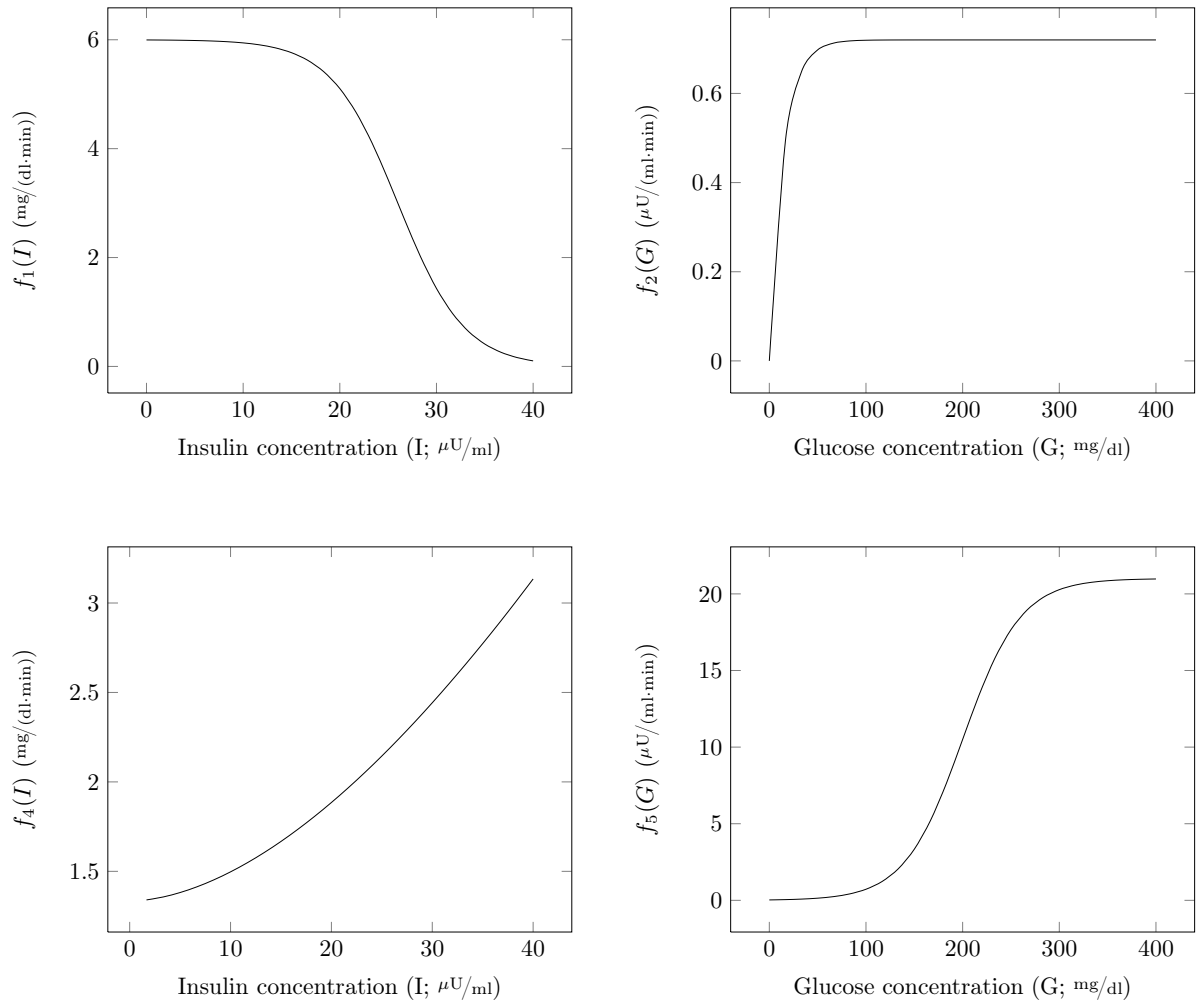
\begin{figure}[htbp]
\begin{centering}
\begin{tikzpicture}[domain=0:40,scale=.9] 
\begin{axis}[ xlabel=Insulin concentration (I; \unitfrac{$\mu$U}{ml}), ylabel=$f_1(I)$ (\unitfrac{mg}{(dl$\cdot$min)}) ] 
\addplot [black, smooth]  {10/3*(180/(1 + exp(0.29*(x*(3/10)*10/3 - 26))))/100};
\end{axis} 
\end{tikzpicture} 
\hspace{.5cm}
\begin{tikzpicture}[domain=0:400,scale=.9]
\begin{axis}[ xlabel=Glucose concentration (G; \unitfrac{mg}{dl}), 	ylabel=$f_2(G)$ (\unitfrac{$\mu$U}{(ml$\cdot$min)}) ]
\addplot [black, smooth]  {(72*(1-exp(-x*100/(144*10))))/100};
\end{axis}
\end{tikzpicture}
\par\end{centering}

\begin{centering}
\vspace{1cm}

\par\end{centering}

\begin{centering}
\begin{tikzpicture}[domain=0:40,scale=.9] 
\begin{axis}[ xlabel=Insulin concentration (I; \unitfrac{$\mu$U}{ml}), ylabel=$f_4(I)$ (\unitfrac{mg}{(dl$\cdot$min)}) ]
\addplot [black, smooth]  {10/3*(40 + (940 - 40)/(1 + exp(-1.77*ln((3/10)*x*10/80*(1/11 + 1/(0.2*100))))))/100}; 
\end{axis}
\end{tikzpicture}
\hspace{.5cm} 
\begin{tikzpicture}[domain=0:400,scale=.9]
\begin{axis}[xlabel=Glucose concentration (G; \unitfrac{mg}{dl}), ylabel=$f_5(G)$ (\unitfrac{$\mu$U}{(ml$\cdot$min)})]
\addplot [black, smooth]  {210/(1 + exp((2000 - x*100/10)/300))/10};
\end{axis}
\end{tikzpicture}\caption{Functional forms of $f_{1}$, $f_{2}$, $f_{4}$, and $f_{5}$, from
\citet{Li2006}}

\par\end{centering}

\centering{}\label{fig:meal_curves} 
\end{figure}

We next consider hepatic glucose production, denoted by $f_{1}(I(t-\tau_{2}))$.
The equation for $f_{1}$ is given in Table \ref{tab:f1f5} and its
shape in Figure \ref{fig:meal_curves}. Insulin inhibits hepatic production,
so it makes sense that $f_{1}$ would be a decreasing function of
insulin concentration. Furthermore, there is a well-documented time
delay between when insulin reaches the liver and when the liver responds
by adjusting glucose production rate (\citet{Li2006}). We denote
this delay as $\tau_{2}$, the amount of time (in minutes) required
for a change in insulin concentration to affect hepatic glucose production.

Glucose concentration can also decrease via two pathways, namely (1)
utilization by the central nervous system and (2) utilization by muscle
and fat cells. Glucose utilization by the central nervous system (CNS)
does not depend on insulin concentration; these cells will use all
of the glucose available to them up to a threshold. We represent this
behavior with $f_{2}$, whose equation is given in Table \ref{tab:f1f5}
and shape in Figure \ref{fig:meal_curves}. Muscle and fat cells,
on the other hand, do rely on the presence of insulin to take up glucose;
thus, we represent their consumption with the product $f_{3}(G(t))\cdot f_{4}(I(t))$.
Here we arrive at the first complication that diabetic illness introduces;
the muscle and fat cells of type II diabetics are less sensitive to
insulin, and thus cannot take up glucose from the blood stream as
easily a non-diabetic person's cells. The scaling factor $\gamma$
accounts for this; $\gamma$ can take values from 0 to 1, with 0 corresponding
to no ability for muscle and fat cells to take up glucose, and 1 corresponding
to a non-diabetic person's glucose uptake ability. Thus, lower values
of $\gamma$ correspond to more severe cases of diabetes. The additional
factor, $[1+s(m-m_{b})]$, accounts for the positive effect of exercise
on insulin sensitivity (\citet{Devlin1992}). Here, $m$ corresponds
to minutes of moderate to vigorous physical activity (MVPA) per day.
60 minutes per day of MVPA ($m_{b}=60$) is considered average; any
less than this decreases glucose tolerance and any more increases
glucose tolerance, with the effect of exercise more significant for
individuals with better baseline glucose tolerance ($\gamma$ close
to 1). Nelson et al.\ observed a decrease in insulin resistance corresponding
to an increase in physical exercise; this data is well-modeled by
the line $y=1.9127-0.0072x$ (\citet{Nelson2013}). The slope of this
line, interpreted as the percent increase in glucose tolerance for
each additional minute of exercise, provides the rationale behind
multiplying by $s=0.0072$ in Equation (\ref{eqn:Gmodel}).

Now let us justify the equation for insulin concentration, Equation
(\ref{eqn:Imodel}). Like glucose, there are two pathways by which
insulin concentration can increase: (1) insulin infusion and (2) pancreatic
$\beta$-cell production. We let the insulin infusion $I_{in}$ be
a constant since periodic insulin infusion would only make sense if
glucose intake were also periodic. We represent pancreatic $\beta$-cell
production with the function $f_{5}(G(t-\tau_{1}))$, defined in Table
\ref{tab:f1f5} and with form given in Figure \ref{fig:meal_curves}.
As illustrated in the schematic diagram, elevated blood glucose encourages
pancreatic insulin production. There is a delay before the pancreas
can respond to changes in blood glucose, for which $\tau_{1}$ accounts
(\citet{Li2006}).

Finally, there is one significant way for insulin concentration to
decrease, which is through metabolism by human insulin-degrading enzyme
(IDE) (\citet{Authier1996}). As an enzymatic reaction, we quantify
insulin degradation with Michaelis-Menten kinetics using the term
$\frac{V_{max}I(t)}{K_{M}+I(t)}$. Here, $V_{max}$ is the maximum
insulin clearance rate and $K_{M}$ is the enzyme's half-saturation
value (\citet{wang2009}).

\begin{table}[t]
\caption{Parameter values for functions $f_{1}-f_{5}$ (from \citet{LiKuangMaxon2006jtb},
\citet{Sturis1991}, and \citet{tolic2000})}

\begin{centering}
\begin{tabular}{|c|l|r|l|}
\hline 
Parameters  & Units  & Values  & Meaning\tabularnewline
\hline 
$\alpha$  & \unitfrac{liter}{mU}  & 0.29  & Scaling factor; sets hepatic sensitivity to changes in insulin\tabularnewline
$\beta$  & $-$  & 1.77  & Scaling factor \tabularnewline
$a_{1}$  & \unitfrac{mg}{liter}  & 300  & Scaling factor; sets pancreatic sensitivity to changes in glucose\tabularnewline
$C_{1}$  & \unitfrac{mg}{liter}  & 2000  & Glucose concentration at which pancreas is most efficient\tabularnewline
$C_{2}$  & \unitfrac{mg}{ liter} & 144  & Scaling factor; sets CNS cell sensitivity to changes in glucose\tabularnewline
$C_{3}$  & \unitfrac{mg}{ liter}  & 1000  & Scaling factor; sets muscle cell sensitivity to changes in glucose\tabularnewline
$C_{4}$  & \unitfrac{mU}{liter}  & 80  & Scaling factor; sets muscle cell sensitivity to changes in insulin\tabularnewline
$C_{5}$  & \unitfrac{mU}{liter}  & 26  & Insulin concentration at which liver is most efficient\tabularnewline
$E$  & \unitfrac{liter}{ min}  & 0.2  & Insulin transport rate from plasma into cells\tabularnewline
$R_{g}$  & \unitfrac{mg}{ min}  & 180  & Maximum hepatic glucose production rate\tabularnewline
$R_{m}$  & \unitfrac{mU}{min}  & 210  & Maximum pancreatic insulin production rate\tabularnewline
$t_{i}$  & \unit{min}  & 100  & Exponential time constant for intercellular insulin degradation\tabularnewline
$U_{0}$  & \unitfrac{mg}{ min}  & 40  & Low-insulin limiting rate of muscular glucose consumption\tabularnewline
$U_{b}$  & \unitfrac{mg}{min}  & 72  & Maximum glucose utilization rate by brain and nerve cells\tabularnewline
$U_{m}$  & \unitfrac{mg}{ min}  & 940  & High-insulin limiting rate of muscular glucose consumption \tabularnewline
$V_{g}$  & \unit{liter}  & 10  & Volume of the body into which glucose can diffuse \tabularnewline
$V_{i}$  & \unit{liter}  & 11  & Intercellular volume\tabularnewline
$V_{p}$  & \unit{liter}  & 3  & Volume of plasma in the body\tabularnewline
\hline 
\end{tabular}
\par\end{centering}

\label{tab:params} 
\end{table}

\section{Analysis\label{sec:Analysis}}

Since type I and type II diabetes differ so significantly in origin
and in the type of therapy required, we will address them separately
here, starting with the more prevalent type II.

\subsection{Type II diabetes}

To lay the groundwork to address our first two objectives, we consider
the fasting case with no insulin therapy, i.e. $G_{in}=I_{in}=0$.
This corresponds to the conditions one would expect for an individual
undergoing diagnostic tests for diabetes, which are normally done
after a fast of at least eight hours (\citet{ADA2013}). We will first
demonstrate how to determine an individual's steady state (or average)
BGC, and will then show how to determine whether that person's glucose
concentration oscillates.

\begin{figure}

\begin{centering}
\includegraphics[scale=0.6]{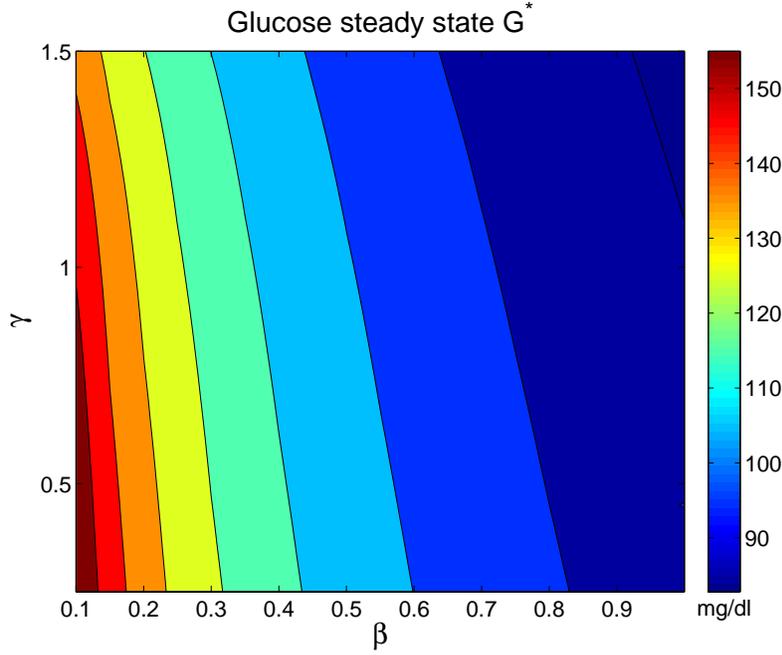} 
\par\end{centering}

\caption{This figure depicts a person's steady state (average) blood glucose
concentration as a function of pancreatic efficiency $\beta$ and
insulin sensitivity $\gamma$. It is apparent, according to the model,
$\beta$ affects blood glucose concentration much more strongly than
$\gamma$. Also, small increases in $\beta$ for a very poorly-functioning
pancreas result in much more dramatic changes in blood glucose concentration
than similar changes for an already well-functioning pancreas. Other
model parameters are held fixed at: $\tau_{1}=5,$$\tau_{2}=15,$
$m=60$, $V_{max}=150$, and $K_{M}=2300$.}

\label{fig:betass} 
\end{figure}

In order to estimate the patient's average BGC we calculate the system's
steady state, setting $G'=I'=0,$ $G(t)=G(t-\tau_{1})=G^{*}$, and
$I(t)=I(t-\tau_{2})=I^{*}$. The constants $G^{*}$ and $I^{*}$ are
the glucose and insulin steady states, respectively; all that remains
is to solve for them. Since $G$ and $I$ arise in functions $f_{1}-f_{5}$
as exponents, bases of exponents, and linear terms, it is very difficult
to solve for the steady states $G^{*}$ and $I^{*}$ analytically.
We instead do so numerically using the default Trust Region algorithm
implemented in \textsc{Matlab}'s \textbf{fsolve} function. Figure
\ref{fig:betass} depicts solutions of the glucose steady state for
varying $\beta$ and $\gamma$, with all other parameters ($\tau_{1},\tau_{2},V_{max},K_{M},m$)
held constant. It is immediately apparent from Figure \ref{fig:betass}
that pancreatic efficiency $\beta$ is much more important than insulin
sensitivity $\gamma$ for determining a patient's average BGC. As
expected, increasing either $\beta$ or $\gamma$ will lead to a lower
BGC.

Now we would like to determine when a person's glucose concentration
will oscillate. To do so, we linearize the model with respect to the
substrates ($G$ and $I$) about the steady state, and then find which
parameter values yield eigenvalues with real part in the positive
half-plane. The linearization gives three Jacobian matrices: 

\begin{center}
$J_{0}=\left[\begin{array}{cc}
-f_{2}'(G^{*})-\gamma[1+0.0072(m-60)]f_{3}'(G^{*})f_{4}(I^{*}) & -\gamma[1+0.0072(m-60)]f_{3}(G^{*})f_{4}'(I^{*})\\
0 & \frac{V_{max}K_{M}}{(K_{M}+I^{*})^{2}}
\end{array}\right]$ 
\par\end{center}

\begin{center}
$J_{1}=\left[\begin{array}{cc}
0 & 0\\
\beta f_{5}'(G^{*}) & 0
\end{array}\right]$ 
\par\end{center}

\begin{center}
$J_{2}=\left[\begin{array}{cc}
0 & f_{1}'(I^{*})\\
0 & 0
\end{array}\right]$. 
\par\end{center}

\begin{figure}

\begin{centering}
\includegraphics[scale=0.6]{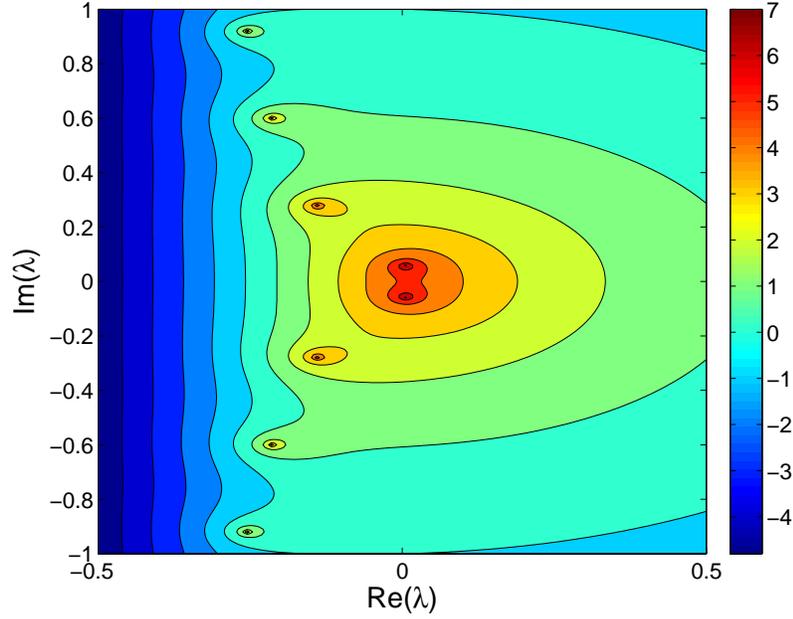} 
\par\end{centering}

\caption{Contour plot of the logged-inverse of the determinant given by the
left-hand side of Equation \ref{eq:det}. The poles indicate the eigenvalues
of the linearized system. These eigenvalues correspond to a non-diabetic
individual, with $\tau_{1}=5$, $\tau_{2}=15$, $\beta=1$, $\gamma=1$,
$m=60$, $V_{max}=150$, and $K_{M}=2300$. Note that the dominant
(rightmost) eigenvalues lie in the positive half-plane, indicating
stable oscillations in blood glucose concentration. \label{fig:eigplot_wide} }
\end{figure}

\begin{figure}
\begin{centering}
\includegraphics[scale=0.6]{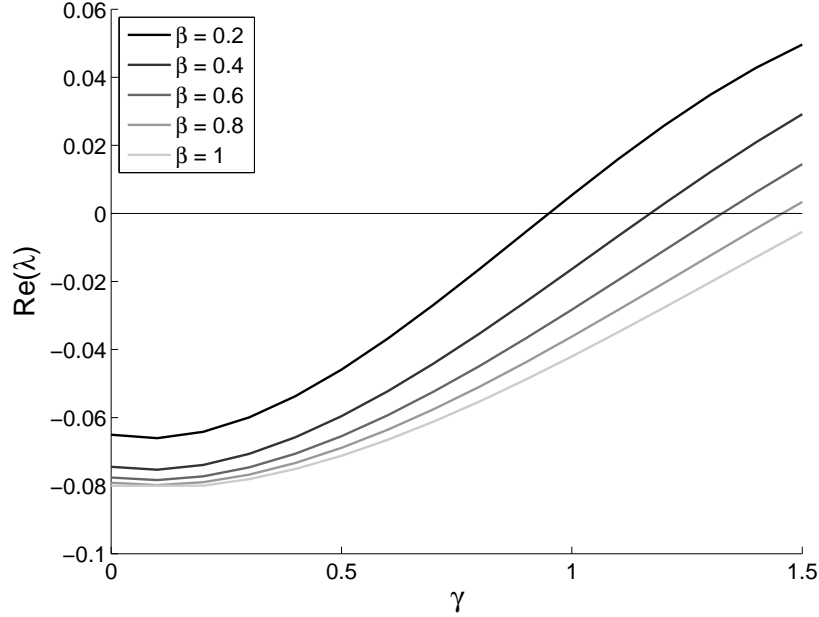} 
\par\end{centering}

\caption{This plot indicates how the real part of the linearized system's dominant
eigenvalue changes with insulin sensitivity $\gamma$ for a few fixed
values of $\beta$. The region below the $Re(\lambda)=0$ line corresponds
to a stable solution, while the region above corresponds to oscillatory
solutions. It appears that higher values of both $\beta$ and $\gamma$
will tend to give rise to oscillatory solutions. \label{fig:eiglines2} }
\end{figure}

The linear system is then 
\[
\left(\begin{array}{c}
G(t)\\
I(t)
\end{array}\right)'=J_{0}\left(\begin{array}{c}
G(t)\\
I(t)
\end{array}\right)+J_{1}\left(\begin{array}{c}
G(t-\tau_{1})\\
I(t-\tau_{1})
\end{array}\right)+J_{2}\left(\begin{array}{c}
G(t-\tau_{2})\\
I(t-\tau_{2})
\end{array}\right)
\]
Assuming solutions of the form $e^{\lambda t}$, we arrive at the
following eigenvalue equation:

\begin{equation}
|J_{0}+e^{-\lambda\tau_{1}}J_{1}+e^{-\lambda\tau_{2}}J_{2}-\lambda{\bf I}|=0\label{eq:det}
\end{equation}
Plotting the inverse of this determinant and looking for poles allows
us to identify which complex values of $\lambda$ solve this equation
and thus are eigenvalues. Figure \ref{fig:eigplot_wide} provides
an example contour plot of this determinant-inverse using parameter
values for a non-diabetic individual. Note that, as expected, the
dominant (rightmost) eigenvalues lie in the positive half plane, indicating
stable blood glucose oscillations. The eigenvalues move as the model
parameters change, with increasing values of $\beta$ and decreasing
values of $\gamma$ tending to induce a leftward shift in the eigenvalues.
The dominant eigenvalues eventually cross the imaginary axis into
the negative real half-plane, indicating the disappearance of ultradian
glucose oscillations. Figure \ref{fig:eiglines2} depicts just how
the real part of the dominant eigenvalue changes for various pancreatic
efficiencies ($\beta$) as a function of insulin sensitivity ($\gamma$).

We are now in a position to determine which parameter values give
an overall healthy blood glucose profile, marked by oscillatory glucose
oscillations in a moderate (80-120 $\unitfrac{mg}{dl}$) range. In
order to simplify our analysis, we note that we can combine the information
given by $\gamma$ and $m$ into a single parameter $\kappa\equiv\gamma[1+0.0072(m-60)]$
that describes a person's overall glucose uptake efficiency. The lowest
physiologically possible value for $\kappa$ occurs when $\gamma=m=0$,
for which $\kappa=0$. A feasible upper value for $\kappa$ occurs
when $\gamma=1$ and $m=120$ (that is, two hours of moderate to vigorous
physical activity daily, which seems reasonable for the most active
individuals), giving $\kappa=1.432$. We will assume that $\kappa$
can range between 0 and 1.5.

\begin{figure}

\begin{centering}
\includegraphics[scale=0.5]{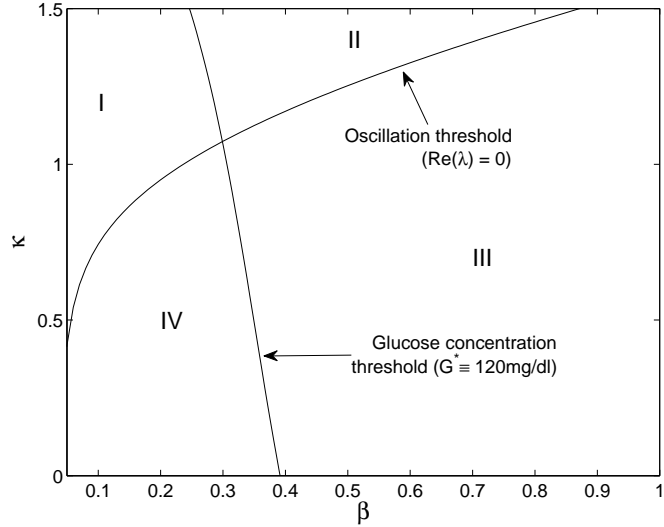} 
\par\end{centering}

\caption{Separation of the parameter space under fasting conditions ($G_{in}=0$)
yielding (I) stable, hyperglycemic solutions; (II) stable, euglycemic
solutions; (III) oscillatory, euglycemic conditions; and (IV) oscillatory,
hyperglycemic conditions. The oscillation threshold line corresponds
to the $\beta$ and $\gamma$ values that make the real part of the
dominant eigenvalue equal to zero. The glucose concentration threshold
line corresponds to the $\beta$ and $\gamma$ values that hold blood
glucose concentration fixed at the upper threshold of 120$\unitfrac{mg}{dl}$.}

\label{fig:fasting_regions} 
\end{figure}

First, we would like to determine which values of $\beta$ and $\kappa$
give steady state solutions between 80 and 120 $\unitfrac{mg}{dl}$.
To do so, we note that, given the steady state values $G^{*}$ and
$I^{*}$, we can solve for the parameters $\kappa$ and $\beta$:
\begin{align}
\kappa= & \frac{-f_{2}(G^{*})+f_{5}(I^{*})}{f_{3}(G^{*})f_{4}(I^{*})}\label{eq:kappa}\\
\beta= & \frac{V_{max}I^{*}}{f_{1}(G^{*})(K_{m}+I^{*})}\label{eq:beta}
\end{align}
To find the $\kappa-\beta$ isocline for the upper glucose threshold,
we set $G^{*}=120$ (equal to the highest acceptable BGC) and vary
$I^{*}$ from 0 $\unitfrac{\mu U}{ml}$ to some high value (100 $\unitfrac{\mu U}{ml}$
is sufficient); thus, $\kappa$ and $\beta$ become parametric functions
with respect to $I^{*}$. This isocline is depicted in Figure \ref{fig:fasting_regions}
by the line that separates Regions I and IV from regions II and III
(the ``glucose concentration threshold''). All values of $\beta$
and $\kappa$ in the fasting case yield average glucose concentrations
above 80$\unitfrac{mg}{dl}$, so we do not show a similar curve for
this lower glucose threshold.

Next, we want to determine which $\beta$ and $\kappa$ values yield
oscillatory solutions. We solve numerically for which $\beta$ and
$\kappa$ make the real part of the dominant eigenvalue calculated
in Equation (\ref{eq:det}) equal to zero, indicating a change in
sign in that eigenvalue's real part. The result is depicted in Figure
\ref{fig:fasting_regions}, with the line that separates Regions I
and II from Regions III and IV (the ``oscillation threshold'').

Let us now take a closer look at Figure \ref{fig:fasting_regions}.
In Regions III and IV we can expect blood glucose oscillations, and
in Regions II and III we will observe blood glucose concentrations
in an acceptable range (i.e. $G^{*}<120\unitfrac{mg}{dl}$). So, for
a patient to have a healthy (non-diabetic) glucose and insulin profile,
he or she should have physiological parameters $\kappa$ and $\beta$
that graphically would appear in Region III.%
\footnote{We here omit $\beta$ values less than 0.05. For such low $\beta$,
pancreatic disability is so acute that the clinical diagnosis would
likely be type I diabetes, which we consider separately at the end
of this section. Mathematically, a singularity lies in this parameter
range that makes numerical solutions difficult to identify.%
}

\begin{figure}

\begin{centering}
\includegraphics[scale=0.5]{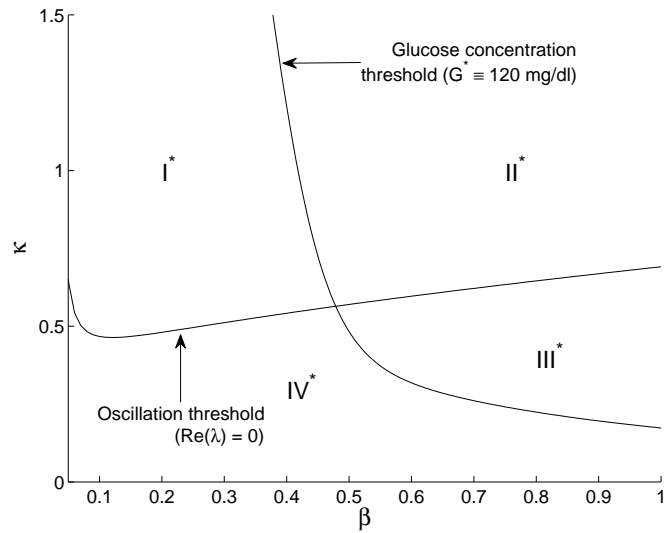} 
\par\end{centering}

\caption{Separation of the parameter space under constant nutrition conditions
($G_{in}=1.08$) yielding (I) oscillatory, hyperglycemic solutions;
(II) oscillatory, euglycemic solutions; (III) stable, euglycemic conditions;
and (IV) stable, hyperglycemic conditions. The oscillation threshold
line corresponds to the $\beta$ and $\gamma$ values that make the
real part of the dominant eigenvalue equal to zero. The glucose concentration
threshold line corresponds to the $\beta$ and $\gamma$ values that
hold blood glucose concentration fixed at the upper threshold of 120$\unitfrac{mg}{dl}$}

\label{fig:nutrition_regions} 
\end{figure}

When we introduce nutrition (external glucose input), we anticipate
needing tighter requirements on the parameters to maintain a healthy
average blood glucose concentration and stable glucose oscillations.
We account for nutrition in the model by setting $G_{in}=1.08\unitfrac{mg}{dl}$
(following \citet{LiKuangMaxon2006jtb}) and repeating the above analysis.
The result is a similar partition of the parameter space, depicted
in Figure \ref{fig:nutrition_regions}. Here, Regions II$^{*}$ and
III$^{*}$ give solutions in which the steady state glucose concentration
is less than $120\unitfrac{mg}{dl}$, and Regions I$^{*}$ and II$^{*}$
give stable glucose oscillations.

\begin{figure}
\begin{centering}
\includegraphics[scale=0.6]{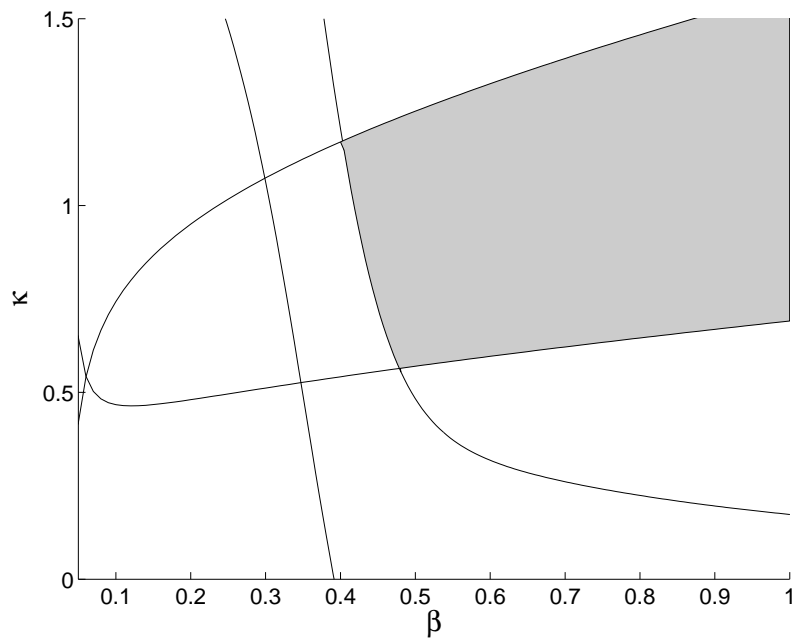} 
\par\end{centering}

\caption{Overlay of fasting and constant-nutrition region plots (Figures \ref{fig:fasting_regions}
and \ref{fig:nutrition_regions}). The shaded area indicates parameter
values that yield oscillatory glucose concentrations in a healthy
range for both fasting and constant nutrition.}

\label{fig:region_overlay} 
\end{figure}

Ideally, we would like to know which physiological parameter values
will give a person a healthy blood glucose profile in both the fasting
and exogenous glucose input cases. To illustrate where these parameters
should lie, we can overlay the plots in Figures \ref{fig:fasting_regions}
and \ref{fig:nutrition_regions} and mark the region for which both
the fasting and the nutrition circumstances predict stable oscillatory
glucose concentrations below $120\unitfrac{mg}{dl}$. This information
is depicted by the shaded region in Figure \ref{fig:region_overlay}.
We observe that if a patient's pancreatic efficiency $\beta$, insulin
sensitivity $\gamma$, and physical activity $m$ can be adjusted
through medication and exercise so that they lie in this region, the
patient's diabetes will be sufficiently controlled. We also note that
this region fills only a portion of the larger region that would give
rise to healthy average blood glucose concentrations while ignoring
ultradian oscillations; that is, simply reducing a diabetic person's
blood glucose levels to a normal range, as is the goal of current
treatment strategies, may not be enough to induce the oscillations
that are characteristic of a healthy blood-glucose profile.

\begin{figure}
\begin{centering}
\includegraphics[scale=0.6]{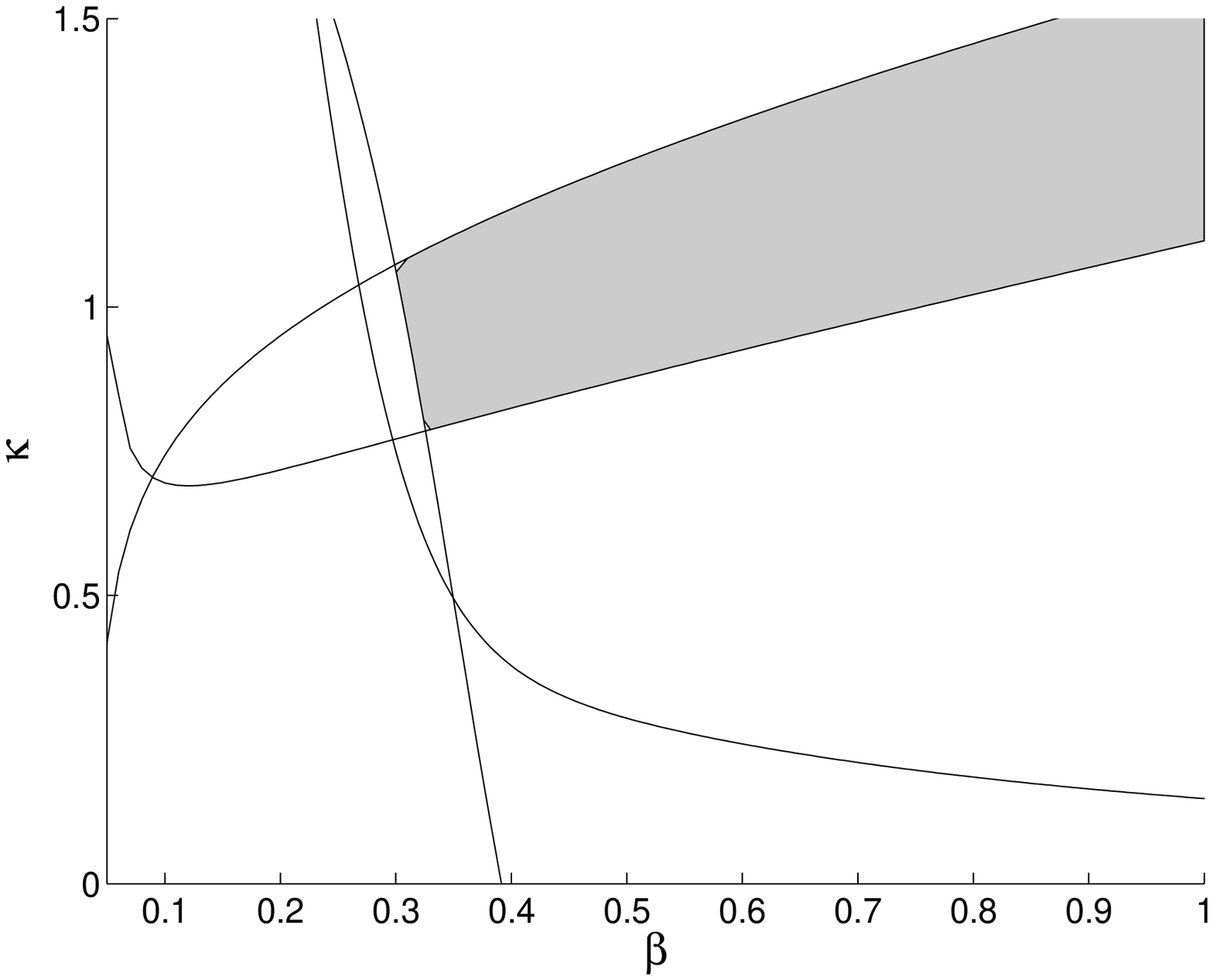} 
\par\end{centering}

\caption{Overlay of fasting and constant-nutrition region plots with insulin
therapy ($I_{in}=0.2$ $\unitfrac{\mu U}{ml\cdot min}$). The shaded
area indicates parameter values that yield oscillatory glucose concentrations
in a healthy range for both fasting and constant nutrition. }

\label{fig:ins_overlay} 
\end{figure}

For one more illustration, let us show how insulin therapy might assist
a patient's treatment strategy. We might expect that insulin therapy
will reduce the need for a well-functioning pancreas (i.e. $\beta$
can be lower), but the effect on insulin sensitivity ($\kappa$) is
more difficult to predict. To address this point, we introduce constant
insulin infusion into the model at a rate of $I_{in}=0.2\unitfrac{\mu U}{ml\cdot min}$
and again determine which $\beta$ and $\kappa$ values yield oscillatory
glucose concentrations in a healthy range under fasting and constant
nutrition. We depict these results in Figure \ref{fig:ins_overlay}.
As predicted, we see that incorporating insulin therapy makes possible
a healthy glucose profile at lower $\beta$ values. However, the region
shrinks in the $\kappa$-direction, suggesting that with insulin therapy
a patient will need to maintain even tighter control over their insulin
sensitivity through some combination of exercise and medication.

\subsection{Type I diabetes}

In the case of type I diabetes, the pancreas is incapable of producing
insulin (i.e.\ $\beta=0$), and so healthy glucose levels can only
be maintained through the injection of external insulin. It is not
possible to induce stable glucose oscillations under these conditions,
but we can still determine how much insulin is required to keep glucose
within a range of 80-120$\unitfrac{mg}{dl}$. Starting with the steady
state relation

\begin{align*}
0 & =G_{in}-f_{2}(G^{*})-\kappa f_{3}(G^{*})f_{4}(I^{*})+f_{5}(I^{*})\\
0 & =I_{in}-\frac{V_{max}I*}{K_{M}+I^{*}}
\end{align*}
we can solve for $\kappa$ and $I_{in}$ as follows:

\begin{align}
\kappa= & \frac{G_{in}-f_{2}(G^{*})+f_{5}(I^{*})}{f_{3}(G^{*})f_{4}(I^{*})}\label{eq:kappa2}\\
I_{in}= & \frac{V_{max}I^{*}}{(K_{M}+I^{*})}\label{eq:I_in}
\end{align}

\begin{figure}
\begin{centering}
\includegraphics[scale=0.6]{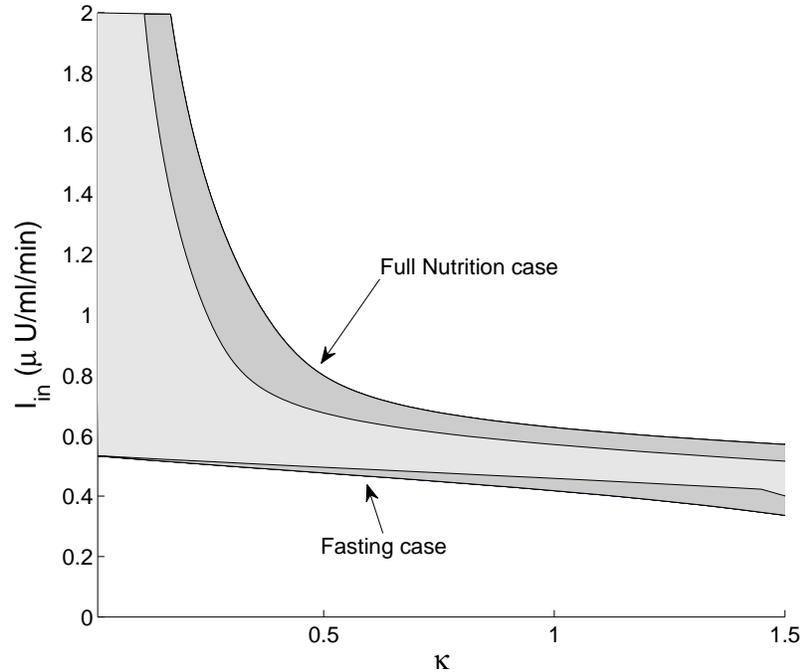} 
\par\end{centering}

\centering{}\caption{This plot indicates how much insulin is required for a type I diabetic
to maintain a healthy BGC given insulin sensitivity $\kappa$. The
dark upper band indicates the insulin infusion rates that will keep
a type 1 diabetic's BGC at an acceptable level with ``full nutrition''
($G_{in}=1.08\unitfrac{mg}{dl\cdot min}$). The dark lower band indicates
the insulin infusion rates necessary to keep a fasting type 1 diabetic's
BGC at an acceptable level ($G_{in}=0$). The lighter region in between
the two bands corresponds to the insulin required to maintain an acceptable
BCG for nutrition levels between fasting and full. \label{fig:type1_regions}}
\end{figure}

As before, we first consider the fasting case, where $G_{in}=0$.
Holding $G^{*}$ fixed at $80\unitfrac{mg}{dl}$ and $120\unitfrac{mg}{dl}$
and treating $I^{*}$ as a parametric variable we can outline a region
in which, for a given $\kappa$, we can find how much insulin ($I_{in}$)
is required to maintain a healthy blood glucose concentration. We
can repeat this process with the ``full nutrition case'' where $G_{in}=1.08\unitfrac{mg}{dl\cdot min}$,
producing a second such region that gives the required insulin when
a person receives nutrition. These regions are depicted in Figure
\ref{fig:type1_regions}. The lighter area between the regions depicts
the insulin infusion rates that would be effective when glucose intake
is somewhere between fasting and full nutrition. As one might expect,
the amount of insulin required increases sharply when the body's insulin
sensitivity ($\kappa)$ becomes low.

\section{Hypothetical case studies \label{sec:Results}}

\subsection{Type II diabetic treatment}

\begin{figure}
\begin{centering}
\includegraphics[scale=0.5]{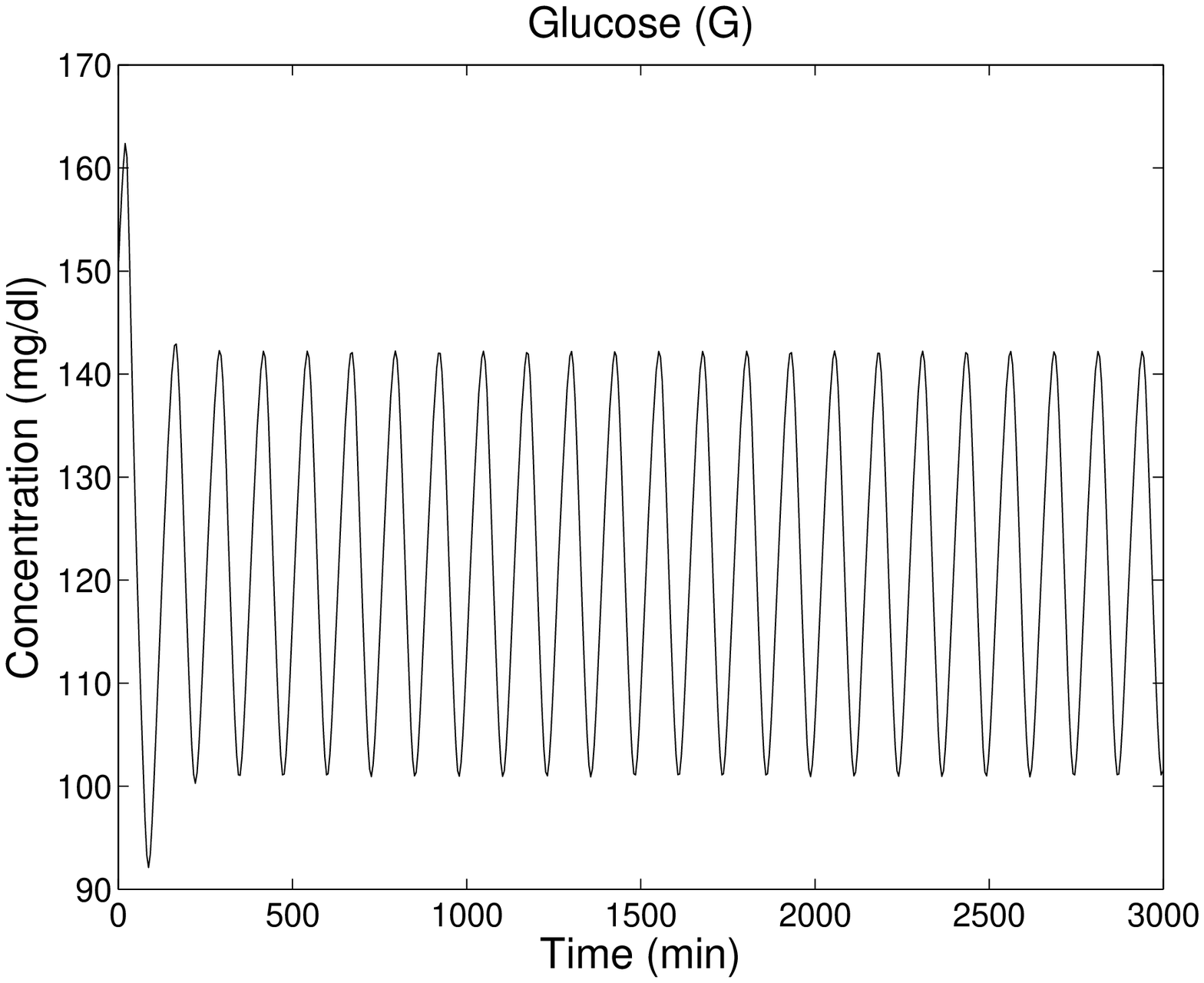}\includegraphics[scale=0.5]{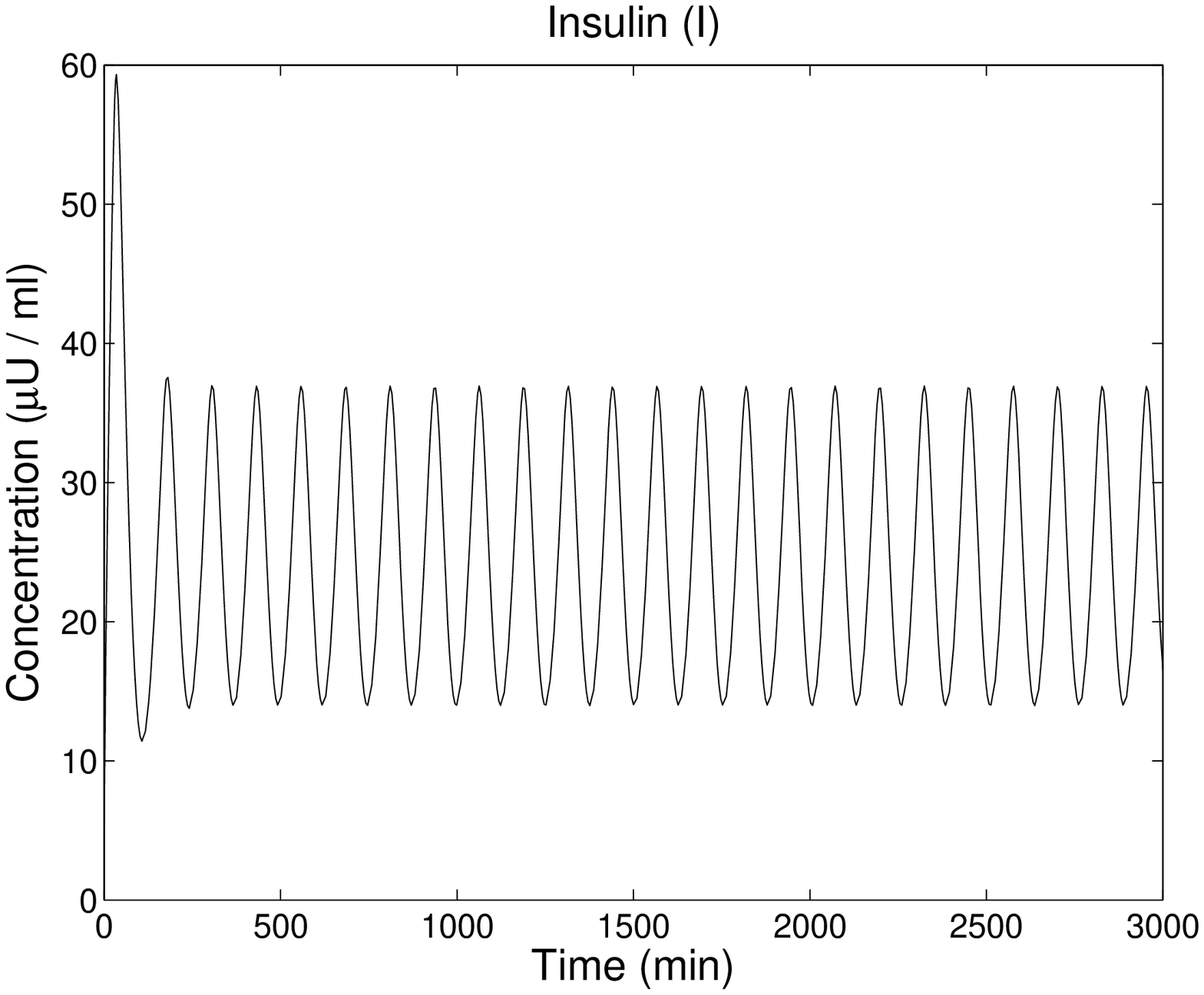}
\par\end{centering}

\begin{centering}
\includegraphics[scale=0.5]{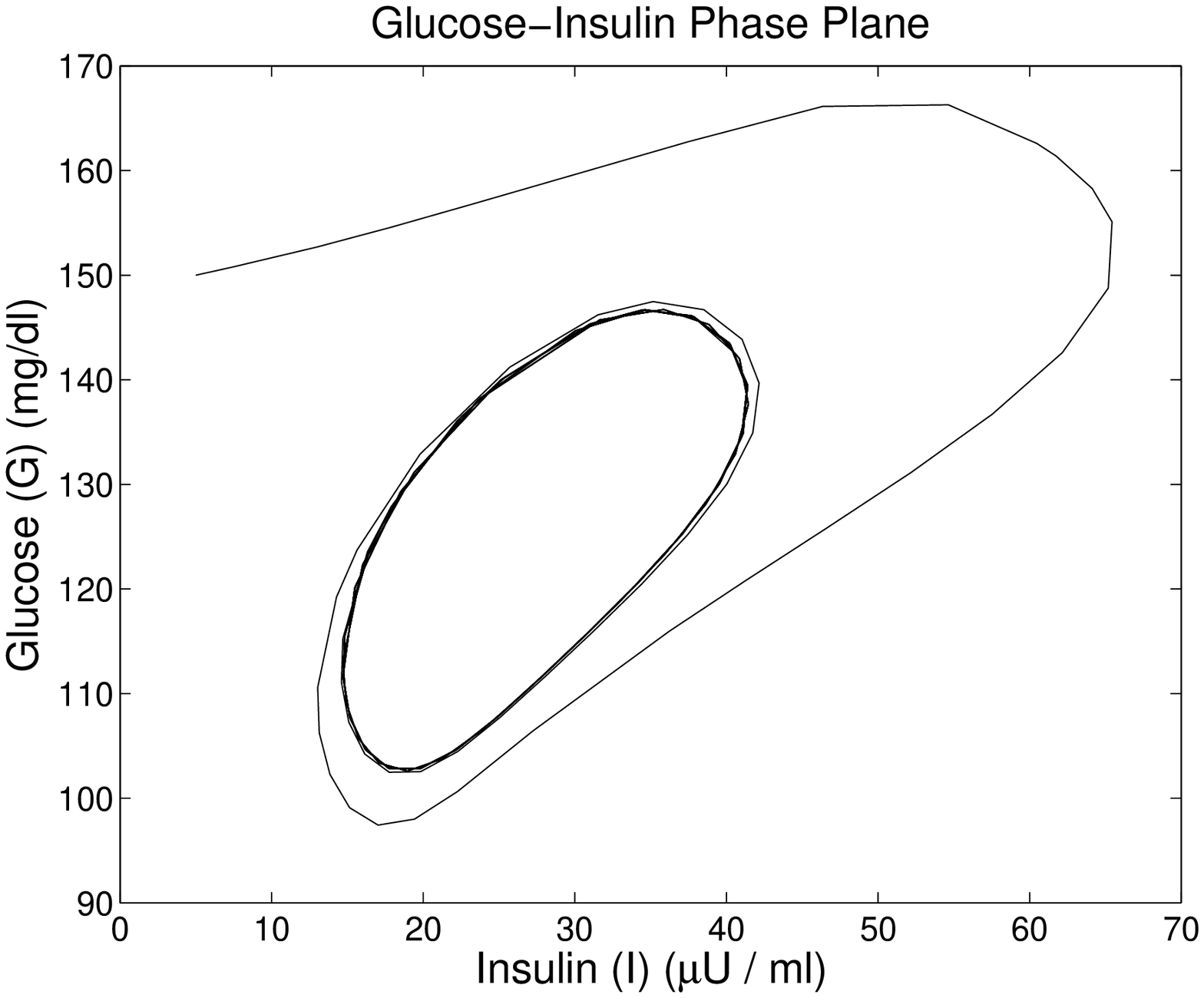}
\par\end{centering}

\caption{Glucose and insulin profiles and phase plane for a type II diabetic
with no treatment ($G_{in}=0$, $I_{in}=0$, $\tau_{1}=5$, $\tau_{2}=15$,
$\beta=.3$, $\gamma=.4$, $m=0$, $V_{max}=150$, $K_{M}=2300$)\label{fig:gi_type2}}
\end{figure}

\begin{figure}
\begin{centering}
\includegraphics[scale=0.5]{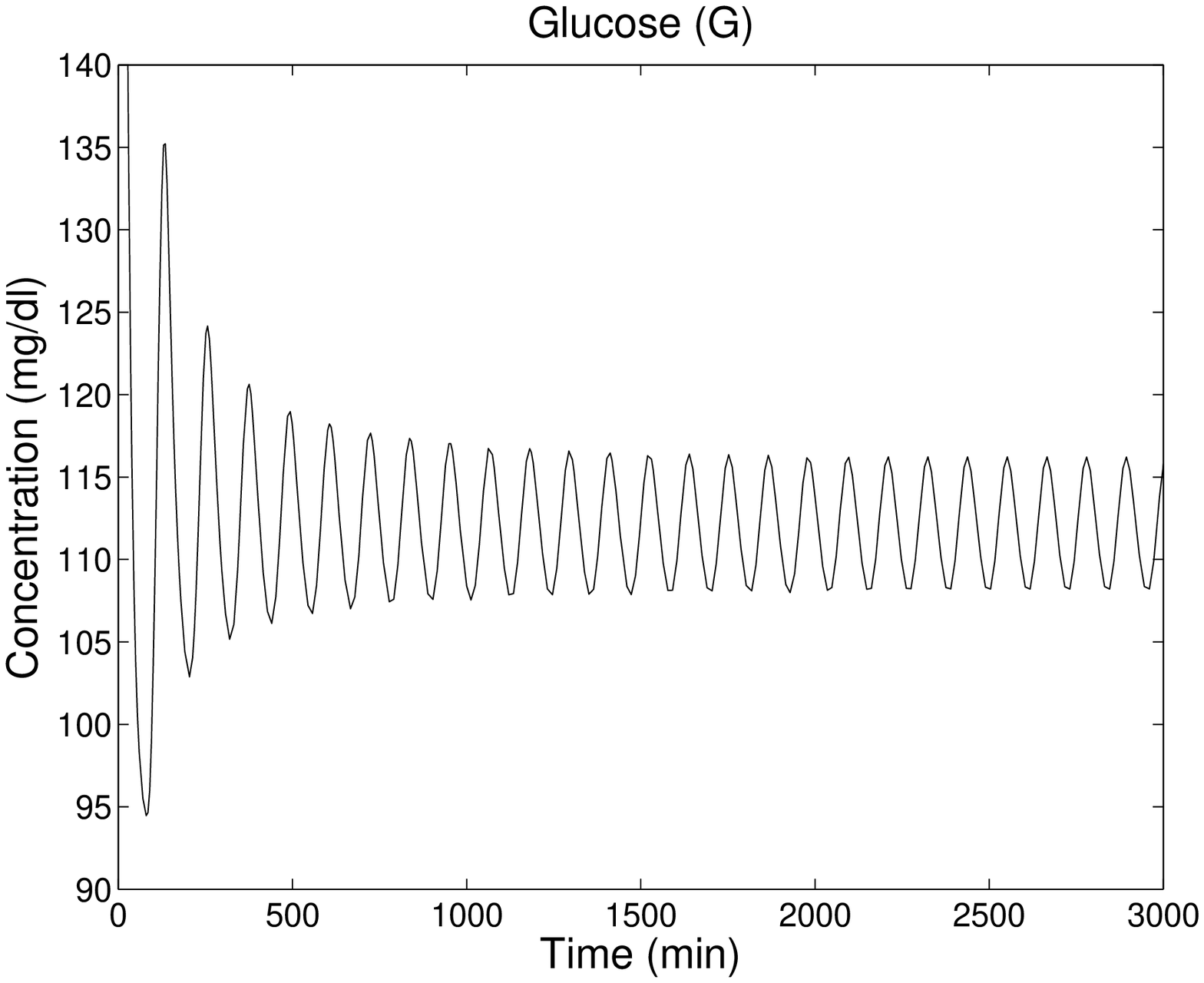}\includegraphics[scale=0.5]{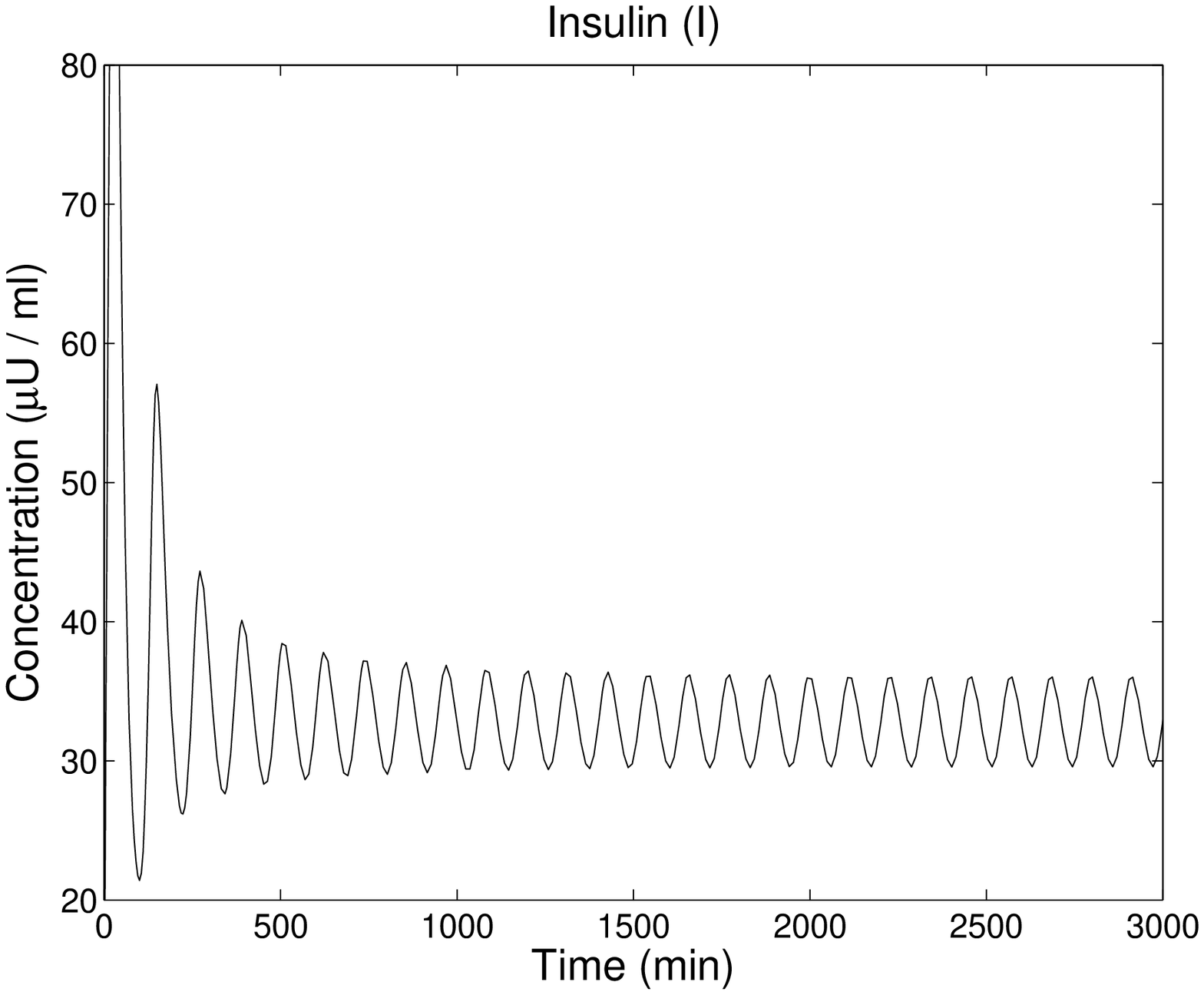}
\par\end{centering}

\caption{Glucose and insulin profiles for a type II diabetic under the first
treatment strategy ($G_{in}=1.08$, $I_{in}=0$, $\tau_{1}=5$, $\tau_{2}=15$,
$\beta=.6$, $\gamma=.7$, $m=60$, $V_{max}=150$, $K_{M}=2300$)\label{fig:type2_t1}}
\end{figure}

\begin{figure}
\begin{centering}
\includegraphics[scale=0.5]{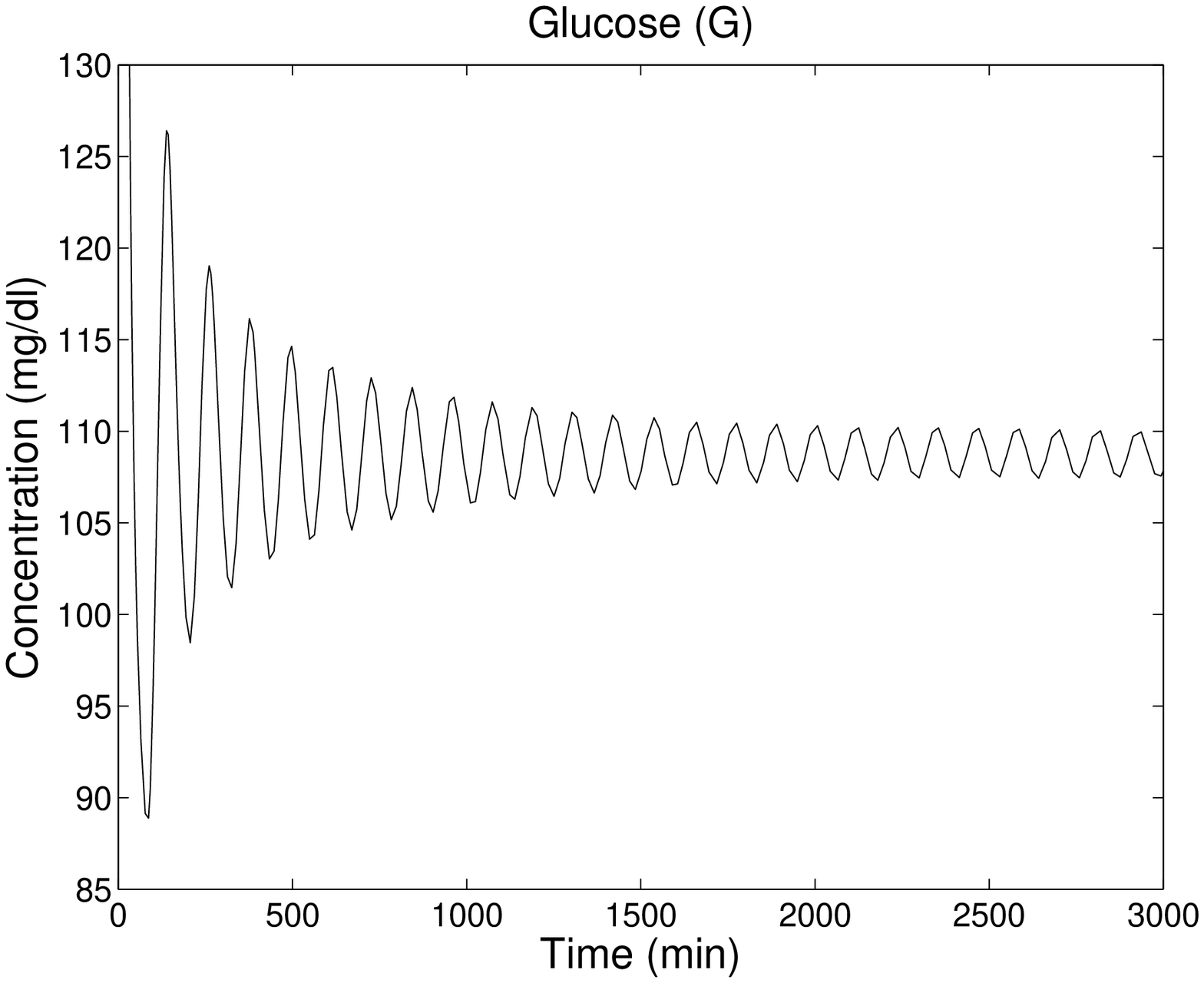}\includegraphics[scale=0.5]{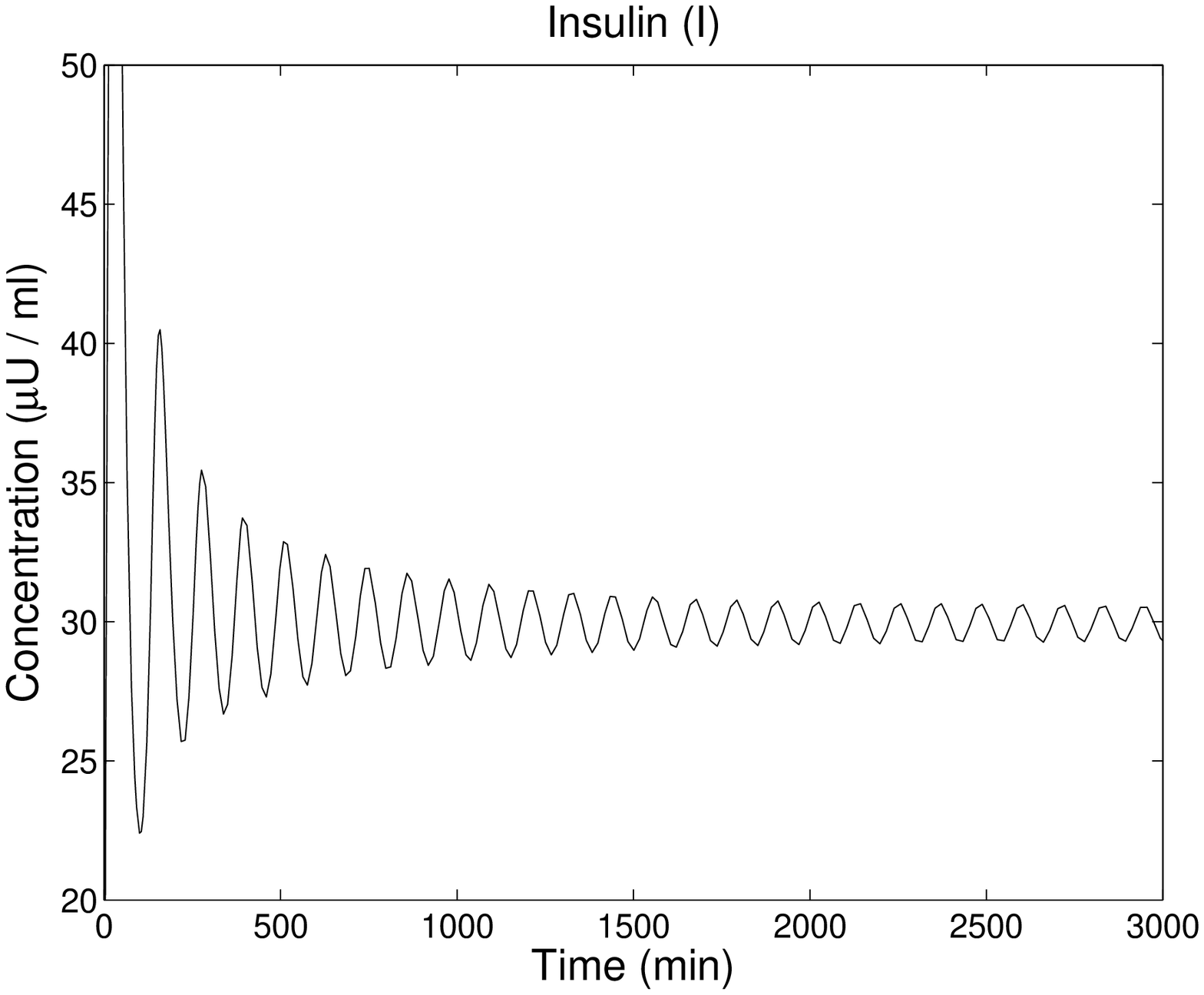}
\par\end{centering}

\caption{Glucose and insulin profiles for a type II diabetic under the second
treatment strategy ($G_{in}=1.08$, $I_{in}=0.2$, $\tau_{1}=5$,
$\tau_{2}=15$, $\beta=.4$, $\gamma=.7$, $m=120$, $V_{max}=150$,
$K_{M}=2300$)\label{fig:type2_t2}}
\end{figure}

To illustrate the value of the analysis in Section \ref{sec:Analysis},
let us consider a hypothetical type II diabetic. The patient's fasting
glucose is measured at 130 $\unitfrac{mg}{dl}$, higher than the ADA's
125 $\unitfrac{mg}{dl}$ threshold for diabetes diagnosis (\citet{ADA2013}).
The patient's doctor measures the patient's pancreatic efficiency
and insulin sensitivity using a euglycemic glucose clamp or an oral
glucose tolerance test and the \emph{minimal model}, two techniques
that have been used in the past to characterize a patient's disease
(\citet{Bergman1989,Brun2012,Ferrannini1998,Stumvoll2000}). Results
show that the patient's pancreas functions at 30\% of normal efficiency
and that the patient's insulin sensitivity is 40\% that of an average
non-diabetic person. The patient lives a largely sedentary life, so
the amount of moderate to physical activity per day the patient receives
($m$) is virtually zero. From this information, the doctor can readily
deduce the parameter values $\beta=0.3$ and $\kappa=0.4\cdot[1+0.0072\cdot(0-60)]=0.23$.
Figure \ref{fig:gi_type2} shows the model's prediction of the patient's
glucose and insulin profile; note that the patient is certainly hyperglycemic,
since the patient's blood glucose concentration regularly exceeds
130 $\unitfrac{mg}{dl}$. By placing the patient's particular $\beta$
and $\kappa$ values on the plot in Figure \ref{fig:region_overlay}
it becomes apparent that the patient requires an increase in both
pancreatic efficiency and insulin sensitivity. There are now several
options that could help re-establish glycemic control. To increase
the patient's insulin sensitivity, the doctor could place the patient
on a medication such as Metformin that would increase $\gamma$ to
0.7, along with introducing 60 minutes of physical activity per day
(\citet{Kirpichnikov2002}). Then, to decrease the patient's BGC,
the doctor could increase the patient's pancreatic efficiency $\beta$
to 0.6 using sulfonylurea drugs (\citet{Aguilar-Bryan1995}). The
model's prediction for this scenario is depicted in Figure \ref{fig:type2_t1}.
Alternatively, the doctor could prescribe insulin therapy in the form
of 0.2 $\unitfrac{\mu U}{ml\cdot min}$ administered continuously
by an artificial pancreas; then, the patient's pancreatic efficiency
would only have to increase to about 0.4 through the use of sulfonylureas.
An additional 60 minutes of daily exercise would increase the patient's
insulin sensitivity enough to give the patient a healthy blood glucose
profile, depicted in Figure \ref{fig:type2_t2}. This example illustrates
how, with proper verification and validation, our analysis and proposed
model could help characterize a specific individual's disease and
inform medical care. Our results make it easy to consider multiple
treatment options involving medication, changes in lifestyle, and/or
medical technology, allowing the patient to choose the lifestyle changes
and therapies that work best him or her.

\subsection{Type I diabetic treatment}

\begin{figure}[htbp]
\centering{}\includegraphics[scale=0.47]{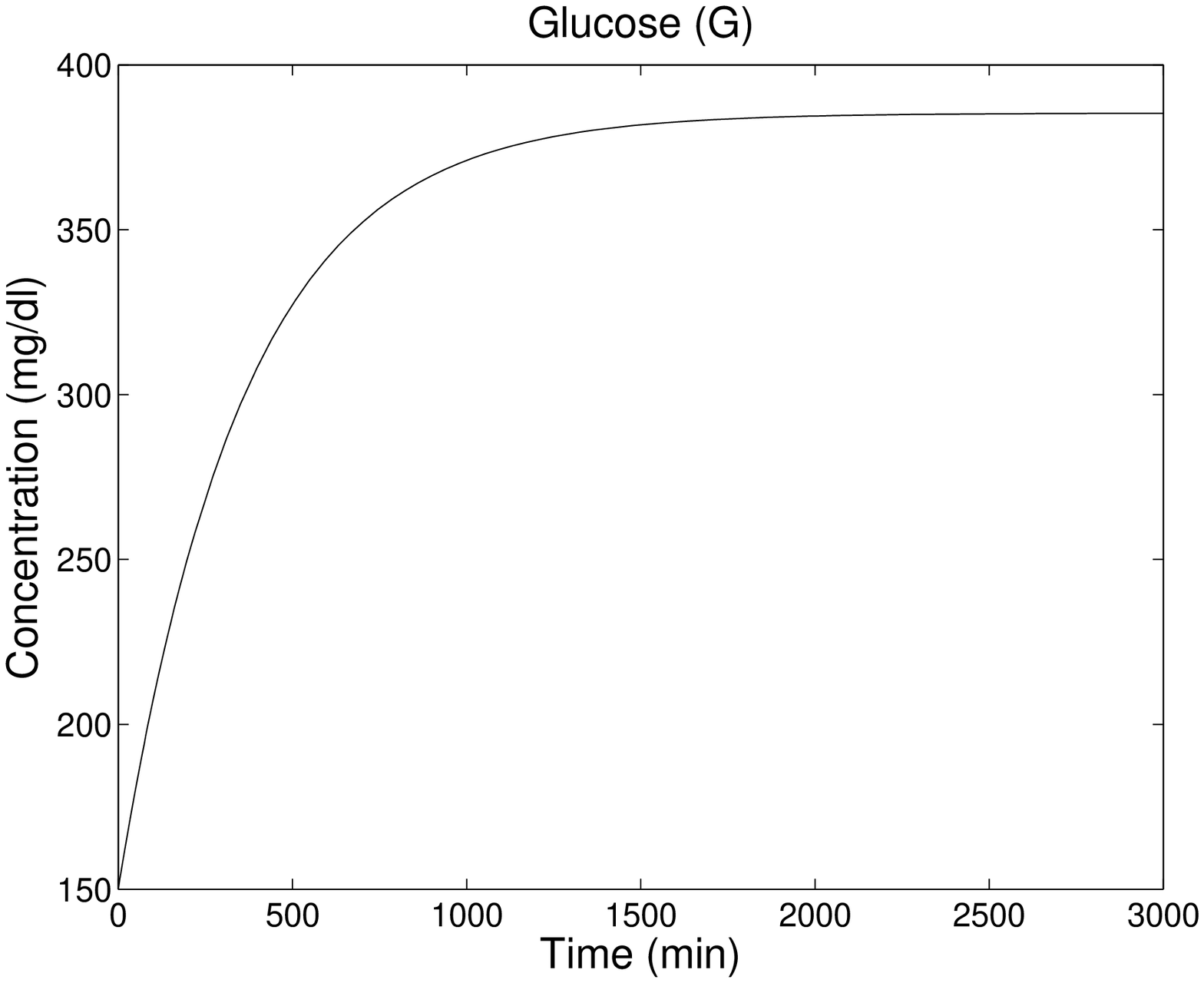}
\includegraphics[scale=0.47]{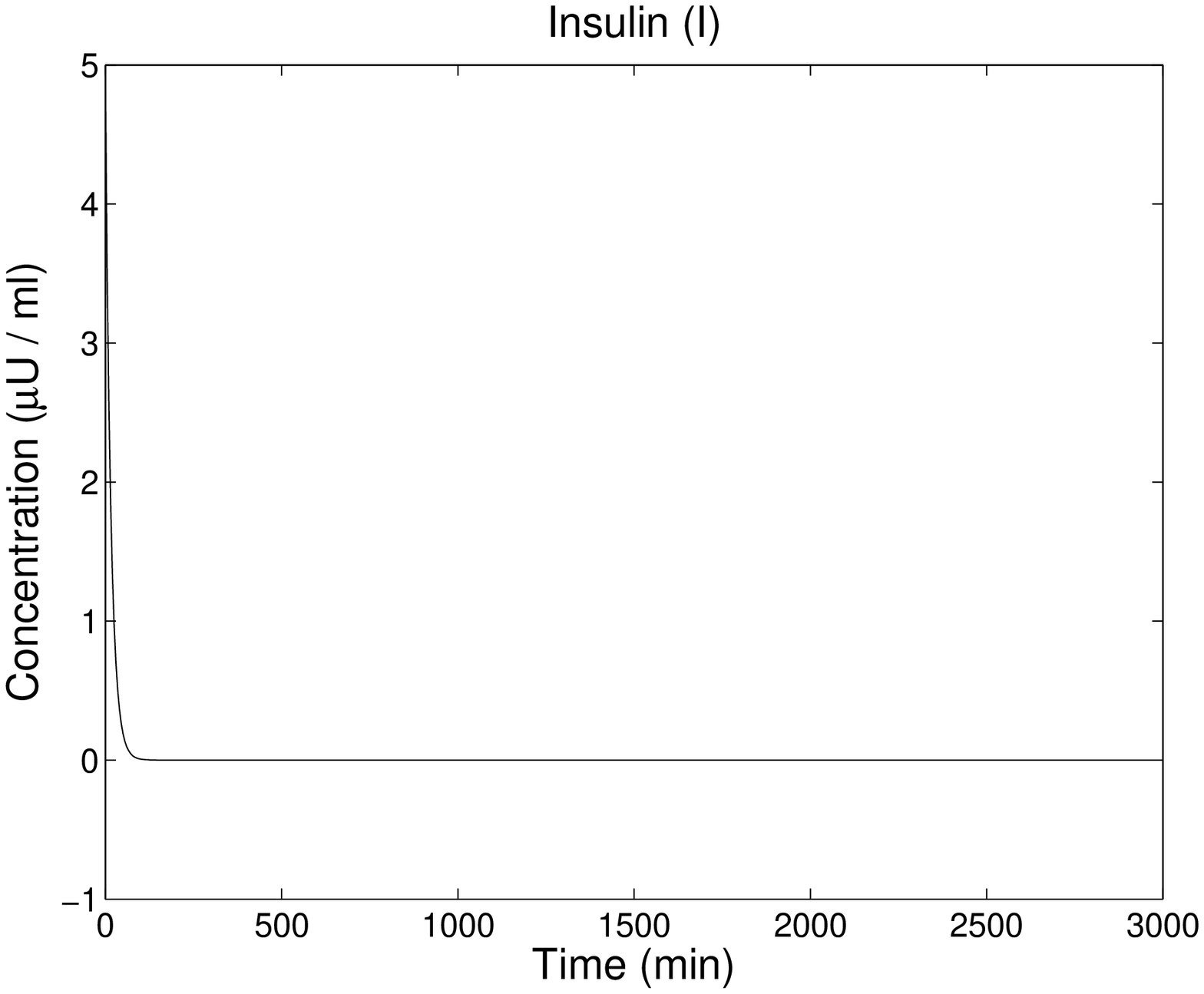} \caption{Glucose and insulin profiles for a type I diabetic with no treatment
($G_{in}=0$, $I_{in}=0$, $\tau_{1}=5$, $\tau_{2}=15$, $\beta=0$,
$\gamma=.75$, $m=60$, $V_{max}=150$, $K_{M}=2300$) \label{fig:type1_notreat}}
\end{figure}

\begin{figure}[htbp]
\centering{}\includegraphics[scale=0.47]{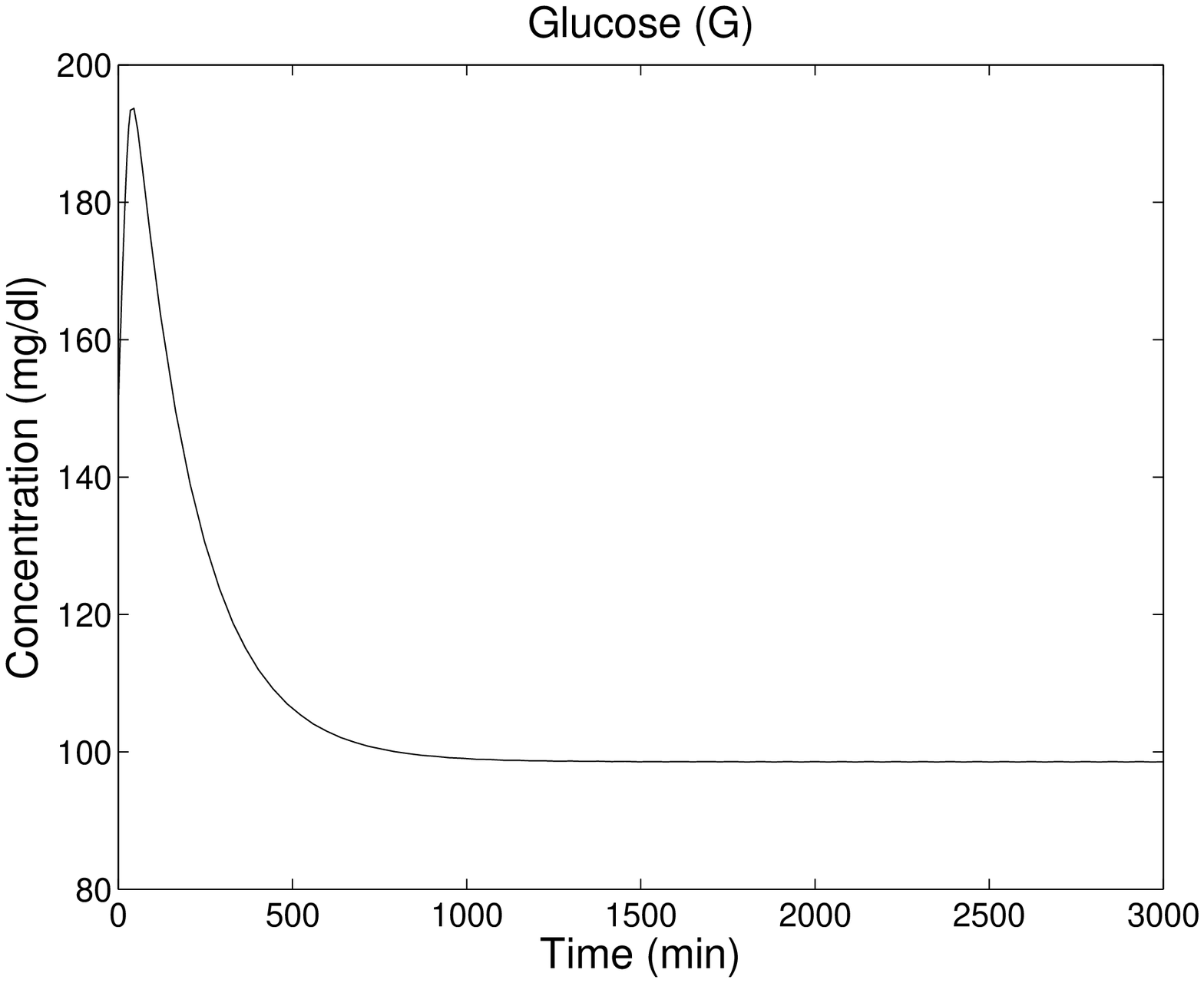}
\includegraphics[scale=0.47]{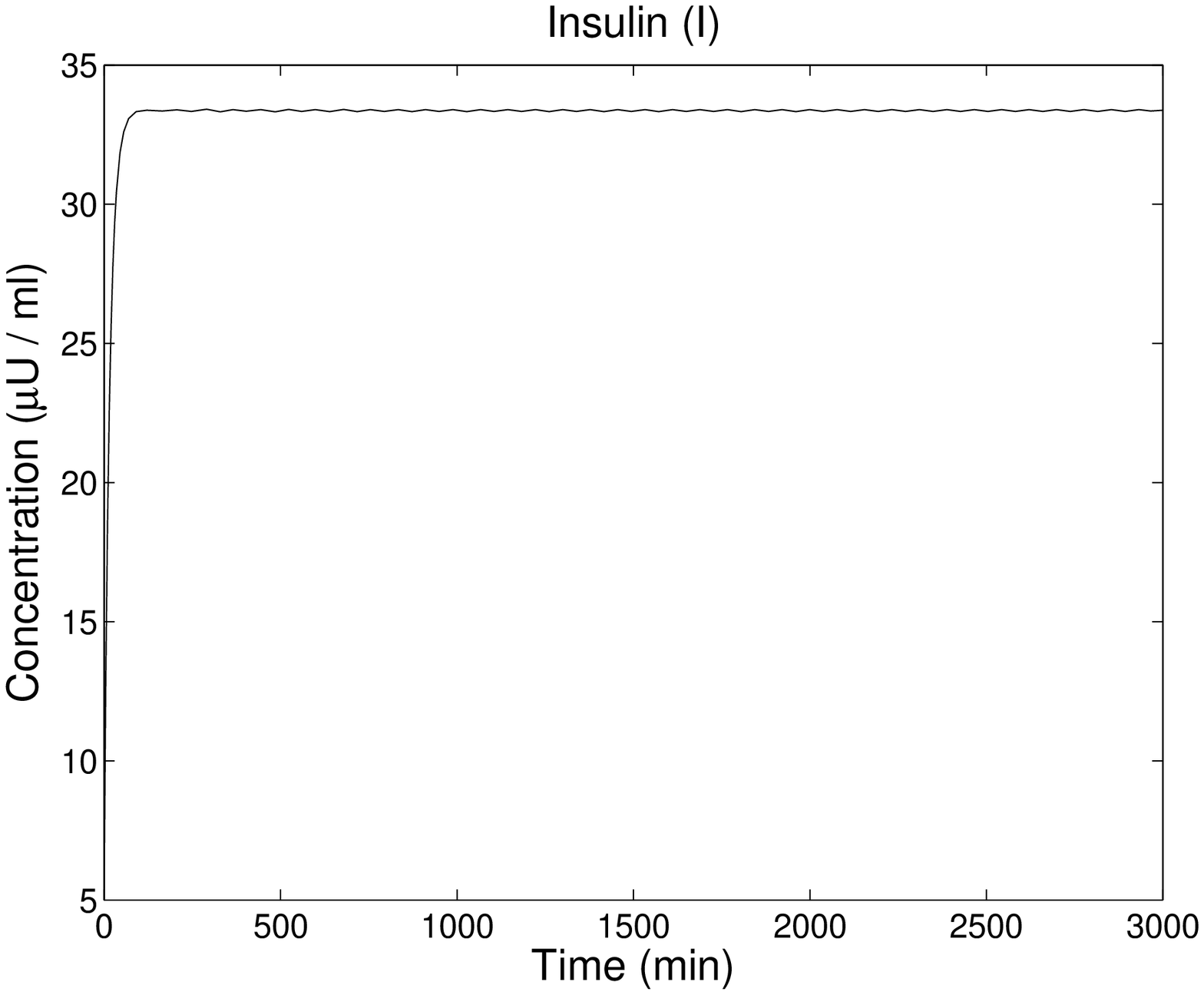} \caption{Glucose and insulin profiles for a type I diabetic with insulin therapy
($G_{in}=1.08$, $I_{in}=0.65$, $\tau_{1}=5$, $\tau_{2}=15$, $\beta=0$,
$\gamma=.75$, $m=60$, $V_{max}=150$, $K_{M}=2300$) \label{fig:type1_treat}}
\end{figure}

Let us now consider a different patient, a type I diabetic ($\beta=0$)
whose insulin sensitivity is about 75\% that of an average non-diabetic
person ($\gamma=.75$), and who receives about an hour of exercise
per day ($m=60$). Figure \ref{fig:type1_notreat} shows this patient's
glucose and insulin profiles in the absence of treatment; with a fasting
steady state BGC greater than three times the ADA-suggested upper
value, an intervention is clearly necessary. Figure \ref{fig:type1_notreat}
also includes a phase portrait of the glucose and insulin concentrations,
demonstrating clear stable oscillations. Similar phase portraits arise
from the remaining glucose-insulin profiles that will be presented,
and are omitted here. To make use of the information depicted in Figure
\ref{fig:type1_regions}, the patient's doctor could first calculate
the patient's particular $\kappa$-value; in this case, $\kappa=0.75[1+0.0072(60-60)]=0.75$.
From Figure \ref{fig:type1_regions}, we see that an insulin infusion
rate between 0.45 and 0.7 $\unitfrac{\mu U}{ml\cdot min}$ from an
artificial pancreas could adequately control the patient's BGC, depending
on the patient's glucose intake rate. Figure \ref{fig:type1_treat}
depicts the patient's glucose and insulin profiles with full nutrition
($G_{in}=1.08$) and an insulin infusion rate of $I_{in}=0.65\unitfrac{\mu U}{ml\cdot min}$;
with a BGC of 100 $\unitfrac{mg}{dl}$, the patient's glucose levels
are adequately controlled.

\section{Conclusions and Future Work\label{sec:Conclusions}}

With this work, we have provided the necessary tools to identify personalized
treatment strategies for diabetic patients based on current clinical
recommendations. We have furthermore shown that common treatment strategies
may omit the ultradian glucose oscillations normally observed in healthy
individuals, and so we lay a framework to ensure that these oscillations
are also maintained. In particular, we have shown that a type II diabetic's
blood glucose levels should be adequately controlled and oscillations
will be maintained when the patient gets an hour of daily exercise
and is placed on a combination of Metformin and sulfonylurea drugs
to increase his or her insulin sensitivity and pancreatic efficiency,
respectively, to 70\% and 60\% of normal. Insulin therapy and an additional
hour of exercise reduce the patient's need for sulfonylureas, requiring
those drugs to increase the patient's pancreatic efficiency to only
40\% of normal. Similarly, we have proposed that a particular type
I diabetic's blood glucose levels can be properly controlled with
a constant insulin infusion between 0.45 and 0.7 $\unitfrac{\mu U}{ml\cdot min}$,
if the patient takes in glucose at a constant rate. With proper verification,
the model presented here could serve as a valuable clinical tool,
helping to provide diabetic patients with a range of treatment options.

This work builds upon a foundation of previous models of the human
glucose and insulin system. Of particular note are those by \citet{Sturis1991},
who presented the foundational model of the system; \citet{Drozdov1995}
who first incorporated a time delay to account for the lag in hepatic
glucose production; and \citet{Li2006} who incorporated a second
time delay to account for the lag in insulin release, and who established
the model most closely associated with the one we present here. \citet{Makroglou2006}
provide a detailed summary of these and other important models in
the field. The work we present marks a shift in focus from much of
what precedes it; the existing literature largely emphasizes the effect
of the two time delays on system's stability, but little has been
done to explicitly analyze the effects of medication and exercise
(\citet{Giang2008jmaa,LiZheng2010mcm,Pei2010}). By explicitly accounting
for these factors, we hope to shrink the gap between theoretical models
and clinical practice, providing individualized information that can
inform clinical care, and to help guide the production of personalized
medical technology. 

Our next steps will involve accounting for non-constant glucose infusion
from meals and for periodic insulin infusion due to injection, in
order to more closely represent the day-to-day variability in a diabetic
person's glucose and insulin intake. We note that, following previous
work in this field, our model does not explicitly account for the
effect of glucagon. While this does not seem to negatively impact
the model's ability to replicate the human glucose regulatory system,
our next models might achieve even better physiological correspondence
by accounting for this additional hormone. We also note that other
organs, such as the kidneys, supplement the liver's glucose production.
Accounting for these organs' effects in future models might yield
further insights into the onset of blood glucose oscillations and
could broaden treatment options. We then plan to develop an algorithm
to determine precisely when and how much insulin should be injected
to maintain a diabetic person's BGC in a healthy range, given their
activity levels, pancreatic efficiency, and insulin sensitivity. We
envision applying this algorithm to an artificial pancreas which,
with a small amount of initial programming and information from an
embedded accelerometer, would require minimal user input. A recent
review of existing devices calls for ``smart control algorithms''
for that better control glucose and insulin oscillations; we anticipate
that this work will respond directly to that call (\citet{Cobelli2011}).
The development of such a device would undoubtedly allow people with
diabetes to live freer, simpler, and healthier lives.

\section{Acknowledgements}

The authors would like to acknowledge the valuable insight given by
Dr.\ Fredric Wondisford at Johns Hopkins University in the initial
stages of this project. We would also like to thank Dr.\ Caroline
Richardson of the University of Michigan for granting permission to
use her team's findings on the effect of exercise on insulin sensitivity.
Finally, we would like to thank the Boettcher Foundation for making
this work possible through their funding.

\bibliographystyle{elsarticle-harv}
\bibliography{mathbioCU}

\end{document}